\documentclass[twocolumn, superscriptaddress, floatfix, prb, aps, showpacs, longbibliography]{revtex4-2}
\usepackage{graphicx, amssymb, color, amsmath}
\usepackage{nicefrac}
\usepackage{multirow, array, booktabs}
\usepackage{float}
\usepackage{mathtools}
\usepackage{bbm}
\usepackage{mathrsfs}
\usepackage{IEEEtrantools}
\usepackage{relsize}
\usepackage{mathrsfs}
\usepackage{mathtools}
\usepackage{xparse}
\usepackage[titletoc, title]{appendix}
\usepackage[colorlinks, bookmarks=true, citecolor=blue, linkcolor=red, urlcolor=blue]{hyperref}
\usepackage{orcidlink}

\graphicspath{{./figures/}}
\begin{document}

\title{Level statistics of the disordered Haldane-Shastry model with $1/r^\alpha$ interaction} 

\author{Vengatesan Ganapathy\orcidlink{0009-0002-9370-587X}}
\email{vengatesanganapathy001@gmail.com}
\affiliation{Institute of Mathematical Sciences, CIT Campus, Chennai 600113, India}
\affiliation{Homi Bhabha National Institute, Training School Complex, Anushakti Nagar, Mumbai 400094, India} 

\author{Pranay Patil}
\email{pranay.patil@iitm.ac.in}
\affiliation{Department of Physics, Indian Institute of Technology Madras, Chennai 600036, India}
\affiliation{Center for Quantum Information, Communication and Computation (CQuICC),Indian Institute of Technology Madras, Chennai 600036, India}

\author{Ajit C. Balram\orcidlink{0000-0002-8087-6015}}
\email{cb.ajit@gmail.com}
\affiliation{Institute of Mathematical Sciences, CIT Campus, Chennai 600113, India}
\affiliation{Homi Bhabha National Institute, Training School Complex, Anushakti Nagar, Mumbai 400094, India}

\date{\today}

\begin{abstract}
Understanding how the interaction range and various types of disorder affect the level statistics of many-body quantum systems, leading to the emergence of many-body localization (MBL), is a challenging open frontier. We study the level statistics of a variant of the spin-$1/2$ Haldane-Shastry model with $1/r^{\alpha}$ interactions, where $\alpha{\geq}0$ parametrizes the range of the interactions, in the presence of position disorder and/or random magnetic fields. We find that neither position disorder nor random magnetic fields alone yields pristine Poisson statistics in this system; however, Poisson statistics emerge in their combined presence, suggesting the emergence of MBL when both types of disorder coexist. Interestingly, once random magnetic fields break the $SU(2)$ symmetry, the strength of the position disorder, $\delta$, appears to play an important role, as evidenced by an approximate scaling collapse of the disorder-averaged gap ratios that is parametrized in terms of a single parameter, $\alpha \delta$.            
\end{abstract}

\maketitle

\section{Introduction}

The various phases that isolated quantum many-body systems can realize have garnered renewed attention in the past few years. The most prominent among these phases is the ergodic one, where for any sizable subsystem, its complement effectively acts like a thermal bath for it. In the ergodic phase, any generic initial state evolves to a featureless thermal state under the system's unitary dynamics at long times and forgets the details of its microscopic initial conditions. This behavior is well-captured by the eigenstate thermalization hypothesis (ETH)~\cite{Deutsch91, Srednicki94, Rigol2008, Polkovnikov_2011, Rigol_2012, Kim_2014, D'Alessio_2016, Deutsch_2018, Ueda2020} which posits that every generic eigenstate is itself thermal. 

ETH can be violated in the presence of disorder that leads to the exponential localization of eigenstates, resulting in the emergence of a non-ergodic phenomenon known as many-mody localization (MBL)~\cite{Anderson58, Lee85, Gornyi_2005, BASKO_2006, Oganesyan_et.al_2007, Huse_2013, Nandkishore2015, Abanin_2019}. In the MBL phase, the subsystem's complement fails to act as a thermal bath, and memory of the initial state is retained for arbitrarily long times. The transition between the ergodic and MBL phases remains an active area of research, and understanding them is central to elucidating how disorder and interactions combine to generate novel non-equilibrium phases of matter. 

A diagnostic for identifying these phases is to compare the spectral statistics of the system to those of the ensembles of random matrix theory (RMT)~\cite{Wigner_1951, Dyson1962, Mehta1990, Haake2018Quantum}. An ergodic phase behaves like a chaotic system, and exhibits level repulsion~\cite{Bohigas84, Haake2018Quantum} and its level statistics are expected to match one of the Gaussian ensembles~\cite{Oganesyan_et.al_2007, Pal_2010}, depending on the system's symmetries. Time-reversal symmetric systems' level statistics match those of the Gaussian orthogonal ensemble (GOE) of real symmetric matrices, while systems lacking time-reversal symmetry, such as those that arise in a magnetic field or are in the presence of magnetic impurities, exhibit level statistics consistent with those of the Gaussian unitary ensemble (GUE) of complex Hermitian matrices. For systems where the time-reversal operator squares to negative identity, such as half-integer spin systems satisfying Kramers' theorem, the spectral statistics match those of the Gaussian symplectic ensemble (GSE) of quaternionic matrices. 

In the MBL phase, the exponential localization of eigenstates prevents energy levels from mixing, resulting in spectral statistics that follow the Poisson distribution~\cite{Oganesyan_et.al_2007, Pal_2010}, which is a hallmark of systems that lack level repulsion. MBL systems can be viewed as being effectively integrable due to the presence of an extensive number of emergent local integrals of motion~\cite{Serbyn_2013, Huse_2014}. 

An efficient and simple way to compare the spectral statistics is via the computation of the statistics of the gap ratios, $\tilde{\mathfrak{r}}_{n}$~\cite{Oganesyan_et.al_2007, Atas_Bogomolny_Giraud_Roux_2013}, which is the ratio of consecutive level spacings, defined as
\begin{equation}
\label{eq: ratio_consecutive_level_spacings}
\tilde{\mathfrak{r}}_n = \frac{\min(d_n, d_{n-1})}{\max(d_n, d_{n-1})} = \min\left(\mathfrak{r}_n, \frac{1}{\mathfrak{r}_n}\right).
\end{equation}
Here the gap between consecutive levels, $d_i{=}E_{i{+}1}{-}E_i$, where $E_{i}$ is the energy of the $i^{\rm th}$ eigenstate, and $\mathfrak{r}_n{=}d_n/d_{n{-}1}$. The average value of $\tilde{\mathfrak{r}}_n$ over the entire spectrum, denoted as $\langle\tilde{\mathfrak{r}}\rangle$, is a single number that captures the spectral statistics and can be readily compared against the corresponding number for the RMT ensembles. The value of $\langle\tilde{\mathfrak{r}}\rangle$ for the RMT ensembles discussed above~\cite{Atas_Bogomolny_Giraud_Roux_2013} are as follows: $\langle\tilde{\mathfrak{r}}\rangle_{\rm Poisson}{=}2\ln(2) {-} 1 {\approx} 0.386$; $\langle\tilde{\mathfrak{r}}\rangle_{\rm GOE}{=}4 {-} 2\sqrt{3} {\approx} 0.536$; $\langle\tilde{\mathfrak{r}}\rangle_{\rm GUE}{=}2\sqrt{3}/\pi {-} 1/2 {\approx} 0.603$; and $\langle\tilde{\mathfrak{r}}\rangle_{\rm GSE}{=}32\sqrt{3}/(15\pi) {-} 1/2 {\approx} 0.676$.

\begin{figure}[htbp!]
    \includegraphics[width=0.5\textwidth]{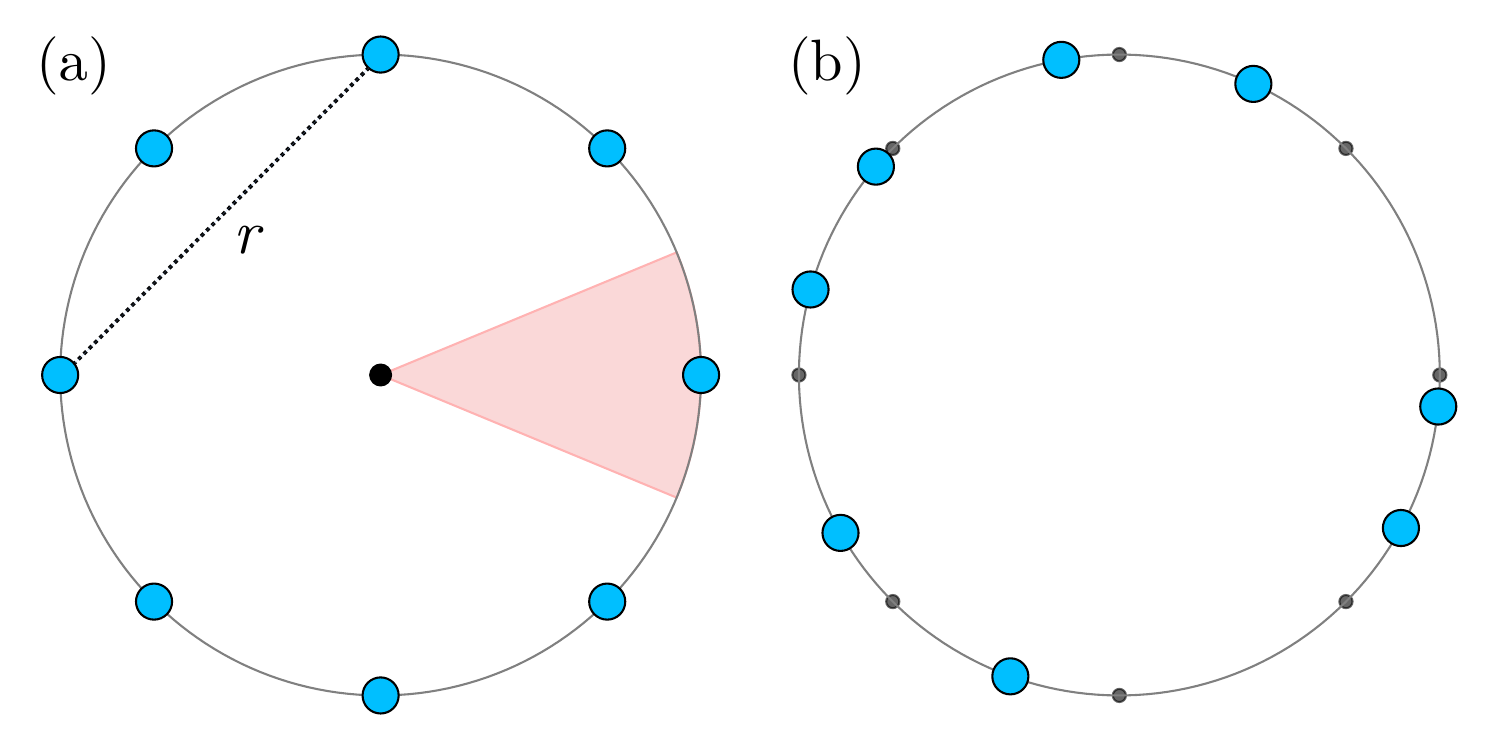}
    \caption{Illustration of the clean lattice (a) and disordered lattice (b) for a realization of the position disorder with disorder strength $\delta{=}1$ for $N{=}8$ sites. Spin-$1/2$ particles are located at the sites marked by blue circles, and the chord distance $r$ is considered as the distance between the spins. The pink sector in the (a) illustrates the angular range in which the rightmost lattice site can lie for the maximum strength of the disorder ($\delta{=}1$).}
    \label{fig: clean_lattice_disordered}
\end{figure}
However, the statistics of gap ratios or level spacing distribution, $P(d)$ with $ d_i{=}E_{i{+}1}{-}E_i$, of the unfolded spectrum \cite{Mehta1990, Kudo_Deguchi_2003} (unfolding smoothens and rescales the spectrum to set the mean level-spacing to unity to enable a fair comparison of the level statistics with those of the RMT ensembles, since the RMT posits an explanation of the correlation between energy levels that is independent of the mean level spacing) only captures the local correlation of the energy levels. Thus, if several uncorrelated matrices with similar statistics of similar sizes get mixed, then the correct underlying statistics can get obscured (see Fig.~\ref{fig: statistics without degeneracies}). This typically happens for quantum many-body Hamiltonians possessing symmetries, which results in a block-diagonal structure for the Hamiltonian. Common symmetries like conservation of total magnetization, parity, and pseudo-momentum (for periodic systems) can be resolved numerically~\cite{Sandvik_2010}, and the Hamiltonian block for particular symmetry sectors can be computed. However, for systems that possess nonlocal symmetries, such as the $sl_2$ loop algebra symmetry emerging in the XXZ model at roots of unity~\cite{FabriciusMcCoy2001}, or the non-local $SU(2)$ symmetry in the spin-1 XY chain~\cite{odea2024levelstatisticsdetectgeneralized}, it may not be readily possible to resolve the symmetries numerically, in which case comparisons with the RMT ensembles can get muddled. These symmetries lead to degeneracies in the spectrum, which makes $\langle\tilde{\mathfrak{r}}\rangle {\to} 0$ (owing to the huge peak at $d{=}0$ in $P(d)$~\cite{Hsu93, Kudo_Deguchi_2003}). Introducing a weak generic perturbation or a small amount of disorder breaks these symmetries, lifting the usually fine-tuned degeneracies, allowing for the spectral statistics of the system to be fairly compared with RMT ensembles.

\begin{figure*}[tbp!]
    \includegraphics[width=1\textwidth]{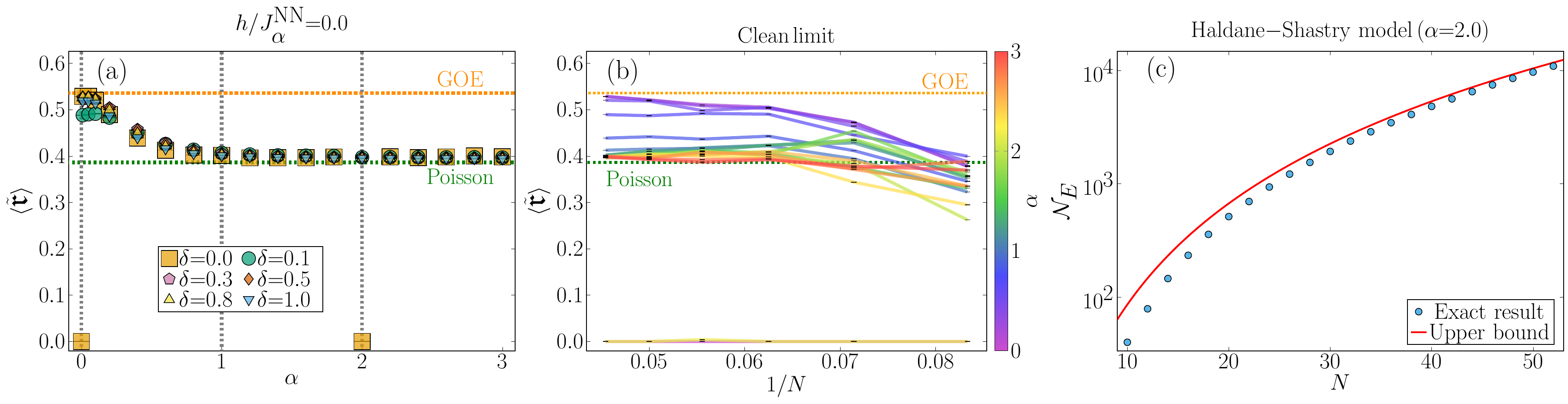}
  \caption{\label{fig: clean_model_plots} (a) Level statistics parameter $\langle \tilde{\mathfrak{r}}\rangle$ of the Hamiltonian of Eq.~\eqref{eq: HS_gen_Hamiltonian} as a function of $\alpha$, where $1/r^{\alpha}$ is the interaction between any two spins separated by a chord distance of $r$, for $N{=}22$ spin-$1/2$ particles on a ring in the $(m,k){=}(0,1)$ symmetry sector, where $m$ is the magnetization and $k$ is the pseudo-momentum [To avoid commutation with parity, results for the second-largest sector with $m{=}0$, i.e., $k{=}1$ are shown (see text).]. The gray vertical dotted lines mark the points at $\alpha{=}0,1,2$. Also shown, are results of disorder averaged level statistics parameter $\langle \tilde{\mathfrak{r}}\rangle$ for the position-disordered Hamiltonian of Eq.~\eqref{eq: z_dis_HS_gen_Hamiltonian}, where $\delta$ parametrizes the strength of the position-disorder, for $N{=}16$ spins in the $m{=}0$ sector; (b) thermodynamic extrapolation of $\langle \tilde{\mathfrak{r}}\rangle$ as a function of $1/N$ for $\alpha {\in}[0,3]$ for the clean system in the $(m,k){=}(0,1)$ sector; (c) number of unique energy states $\mathcal{N}_E$ of the Haldane-Shastry model $\alpha{=}2$ as a function of $N$.}
\end{figure*}

A well-studied example of a system with such symmetries is the Haldane-Shastry (HS) model~\cite{Haldane88, Shastry88}, which is a one dimensional antiferromagnetic Heisenberg model of $S{=}1/2$ spins that are placed uniformly on the unit circle with all-to-all long-range $1/r^{2}$ interactions, where $r$ is the chord distance [see the schematic shown in Fig.~\ref{fig: clean_lattice_disordered}(a)]. Owing to the $SU(2)$ spin-symmetry, the HS Hamiltonian commutes with the total spin, $\vec{S}_{\rm tot}$. Furthermore, the HS Hamiltonian commutes with a so-called rapidity operator $\Lambda$~\cite{Greiter2011}, but $\vec{S}_{\rm tot}$ and $\Lambda$ do not mutually commute. Instead, $\vec{S}_{\rm tot}$ and $\Lambda$ generate an infinite-dimensional associative algebra known as Yangian $Y(sl_2) $~\cite{Yangian_Haldane_et_al_92, Hoft_algebra_DRINFEL, chari1995guide_QG}. This Yangian symmetry of the HS model, along with resulting in its integrability, leads to large degeneracies in its spectrum~\cite{Yangian_Haldane_et_al_92, talstra1995integrability_HS, Squeezed_strings_and_Yangian_Haldane_et_al93, Hoft_algebra_DRINFEL, Talstra_1995}. For an overview of the model, see section 2.2 of Ref.~\cite{Greiter2011}. Despite being integrable, the spectral statistics of the HS model are inconsistent with the Berry-Tabor conjecture~\cite{Berry77} and do not align with any of the RMT ensembles~\cite{Hsu93, Finkel_2005} [see Fig.~\ref{fig: clean_model_plots}(a)]. A natural question to ask is, if we disorder the HS model and lift its degeneracies, would its spectral statistics match any of the RMT ensembles? This is the question we address in this work. 

\begin{figure}[tbp!]
    \centering
    \includegraphics[width=1\linewidth]{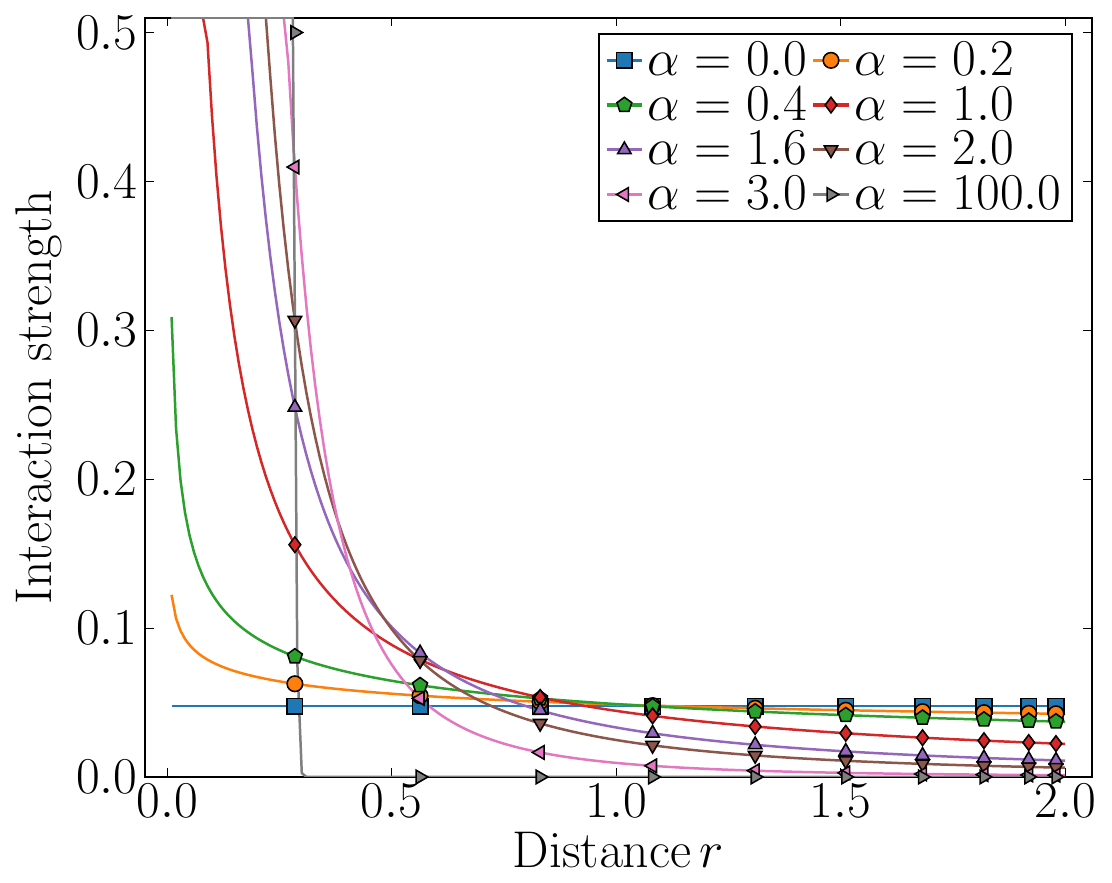}
    \caption{Interaction profile of the Hamiltonian of Eq.~\eqref{eq: HS_gen_Hamiltonian} for $N{=}22$ sites, i.e., the interaction strength between a given site and its neighbors as a function of distance, $r$, for different values of the interaction parameter $\alpha$. The markers indicate the distances from the successive neighboring sites. For the long-range model with $\alpha{<}1$, all interaction strengths are of similar order, while for the short-range model with $\alpha{>}1$, the farthest-neighbor interactions are suppressed. For $\alpha{=}100$, the interaction is almost nearest-neighbor-like since the long-range interactions are heavily suppressed. 
    }
    \label{fig: interaction_profile}
\end{figure}
One way to disorder the HS model is to randomly displace the spins from their roots of unity lattice positions, which introduces randomness in the interactions between the spins~\cite{Cirac_Sierra_2010, Nielsen_2011, Shriya_Pai_2017}. Ref.~\cite{Shriya_Pai_2017} considered this model and showed that the resulting spectral statistics of the disordered HS model still do not align with any of the RMT ensembles. Moreover, the position-disordered HS system does not localize since its $SU(2)$ symmetry is incompatible with area-law scaling of entanglement of its eigenstates, which preclude the formation of a conventional area-law scaling respecting MBL phase~\cite{Vasseur_2015, Potter_2016}.

To further investigate this issue, we consider a generalization of the HS model with $1/r^{\alpha}$ interactions for $\alpha{\ge}0$. We study the spectral statistics of this model in the presence of disorder introduced by random spin positions, longitudinal magnetic field [which breaks the $SU(2)$ spin-symmetry down to $U(1)$], and their combined effects. For the position disordered case, we observe that as $\alpha$ increases from 0, $\langle\tilde{\mathfrak{r}}\rangle$ departs from the GOE value and saturates near 0.399 for $\alpha {\ge} 1$, which is slightly higher than the Poisson value. This behavior is independent of the strength of the position disorder, similar to what Ref.~\cite{Shriya_Pai_2017} found for $\alpha{=}2$. Moreover, neither position disorder nor random magnetic fields alone is sufficient to induce Poisson statistics in this long-range interacting system in the regime where the spin-spin interaction dominates the field strength. Instead, Poisson statistics appear to arise only from their combined effects, suggesting the emergence of MBL in this system when both types of disorder are present. The position disorder strength plays an important role by determining the transition point once the $SU(2)$ symmetry is broken by the random magnetic fields.

The article is organized as follows: In Sec.~\ref{sec: model}, we define the variant of the HS model, with and without disorder, that we consider in this work. Numerical results on the level statistics are presented in Sec.~\ref{sec: results}. We conclude the paper in Sec.~\ref{sec: conclusions} with a summary of our results and an outlook for future directions. 

\section{Model}
\label{sec: model}

We consider the following variant of the HS model described by the Hamiltonian.
\begin{equation}
\label{eq: HS_gen_Hamiltonian}
H_0= \frac{1}{\mathcal{N}}\left( \frac{2\pi }{N}\right)^{\alpha} \sum_{1\leq p<q \leq N} \frac{\vec{S}_p \cdot \vec{S}_q}{\left|\eta_p-\eta_q\right|^\alpha},
\end{equation}
where $\mathcal{N}$ is a scale factor, $\eta_p{=}e^{i\frac{2\pi}{N} p}$ is the position of the $p^{\rm th}$ spin on the unit circle in the complex plane with $i{=}\sqrt{{-}1}$ being the imaginary unit, where $p{=}0,1,{\dots},N{-}1$ label the $N$ roots of unity, where $N$ is always taken to be even. The distance between the sites $p$ and $q$ is taken to be the chord distance $\left|\eta_p{-}\eta_q\right|$ [see Fig.~\ref{fig: clean_lattice_disordered}(a)], and $\alpha{\geq}0$ is a parameter of the model, which controls the interaction strength and range between the spins. The factor $\left( 2\pi /N \right)^{\alpha}$ ensures that the interaction between a pair of spins remains finite in the thermodynamic limit, i.e., as $N {\to} \infty$ [see Eq.~\eqref{eq: max_int_energy}]. When $\alpha{\ll}1$, the interaction is long-ranged, and the strength of the interaction between the farthest pair of spins is comparable to that of the nearest-neighbor spin (see Fig.~\ref{fig: interaction_profile}). However, when $\alpha{\gg}1$, the interaction is short-ranged. As $\alpha$ increases, this model interpolates between the all-to-all Heisenberg model ($\alpha{=}0$) and the XXX or nearest-neighbor Heisenberg model ($\alpha{\to}\infty$), with $\alpha{=}2$ being the Haldane-Shastry point~\cite{Haldane88, Shastry88} (see Fig.~\ref{fig: interaction_profile}). Surprisingly, the ground state in the  $\alpha{>}1$ short-range regime has overlap upwards of $99\%$ with the Jastrow wavefunction that is the exact state of the  $\alpha{=}2$ HS model for all the system sizes considered here (see Appendix~\ref{app: Jastrow}).

An important thing to note is that the energy density of the Hamiltonian of Eq.~\eqref{eq: HS_gen_Hamiltonian}, i.e., the energy per spin, sans the scale factor of $\mathcal{N}$ is extensive and scales as $\mathcal{O}(N^\gamma)$, where $\gamma$ decreases from $1$ to $0$ as $\alpha$ increases from $0$ to $\infty$. To make the energy density intensive, we scale the Hamiltonian of Eq.~\eqref{eq: HS_gen_Hamiltonian} by a factor of $\mathcal{N}{=}\mathcal{O}(N^{\gamma})$, which is commonly referred to as the Kac normalization factor~\cite{Kac1963, Hallam25}. However, since the spectral statistics are composed of the ratio of successive energy differences, they are not affected by the precise value of $\mathcal{N}$.

The position disorder is introduced into the model by randomly displacing the spins from their original sites in such a way that the neighboring spins do not cross each other. This setup is illustrated in Fig.~\ref{fig: clean_lattice_disordered}(b) for a particular disorder realization. The Hamiltonian with position disorder is
\begin{equation}
\label{eq: z_dis_HS_gen_Hamiltonian}
H_\delta= \frac{1}{\mathcal{N}}\left( \frac{2\pi }{N}\right)^{\alpha} \sum_{1\leq p<q \leq N} \frac{\vec{S}_p \cdot \vec{S}_q}{\left|\eta_p^{(\delta)}-\eta_q^{(\delta)}\right|^\alpha},
\end{equation}
where $\eta_p^{(\delta)}{=}e^{i\frac{2\pi}{N}(p{+}\zeta_p^{(\delta)})}$is the new displaced position of the $p^{\rm{th}}$ spin with $\zeta_p^{(\delta)}$ being a uniformly distributed random number in the interval $[{-}\delta/2,\delta/2]$ (illustrated as the pink sector in Fig. \ref{fig: clean_lattice_disordered}(a) for $\delta{=}1$), and $\delta{\in}[0, 1]$ controls the strength of the position disorder. The ground state wave function of the disordered Hamiltonian moves away from the Jastrow form as $\alpha$ deviates from $2$, and even at $\alpha{=}2$ as disorder increases. Nevertheless, even in the presence of disorder, the ground state for $\alpha{>}1$ has $\gtrsim98\%$ overlap with that of the ground state for $\alpha{=}2$ for all the systems considered here, suggesting that the low-energy physics in the entire short-range regime of $\alpha{>}1$ has a commonality (see Appendix~\ref{app: Jastrow}).

The Hamiltonian of Eq.~\eqref{eq: z_dis_HS_gen_Hamiltonian} still preserves the $SU(2)$ spin-symmetry. To break it down to $U(1)$, we next introduce a random longitudinal magnetic field. This is done by adding to Eq.~\eqref{eq: z_dis_HS_gen_Hamiltonian} the term $ H_h{=}(1/\mathcal{N})\sum_{1\leq p \leq N} h_p\, S_p^z $, where $h_p$ is a Gaussian distributed random number with mean $0$ and variance $h^2$, with $h$ being the strength of the random field. Note that in the usual Kac prescription, the interaction term is scaled by the Kac factor $\mathcal{N}$, but the external magnetic field term is not, since if we scale the magnetic field by $\mathcal{N}$, then for large $N$ it makes a negligible contribution to the energy density for $\alpha{>}0$. Here, our focus is on the level statistics obtained from gap ratios, which do not depend on the overall scale. Since we do not know what the exact value of $\mathcal{N}$ is, if we scale the interaction term differently from the magnetic field term, then the level-statistics ratio will be affected by the precise value of $\mathcal{N}$, which is why we have also scaled the external magnetic field term by $\mathcal{N}$. The actual magnetic field should be viewed as being $\mathcal{N} h_{p}$, or equivalently, the $h$ above is actually $\mathcal{N}h$. The Hamiltonian with both position and longitudinal magnetic field disorders is
\begin{eqnarray}
\label{eq: z_h_dis_HS_gen_Hamiltonian}
H&=&H_\delta+H_h \\
&=& \frac{1}{\mathcal{N}}\left[\left( \frac{2\pi }{N}\right)^{\alpha} \sum_{1\leq p<q \leq N} \frac{\vec{S}_p \cdot \vec{S}_q}{\left|\eta_p^{(\delta)}-\eta_q^{(\delta)}\right|^\alpha} + \sum_{1\leq p \leq N} h_p \, S_p^z\right]. \nonumber
\end{eqnarray}
A discussion is in order on what energy scale the local fields $\{h_j\}$ should be compared against. We measure the field strength relative to an $\mathcal{O}(1)$ energy scale, which we take to be the interaction energy between nearest-neighbor (NN) spins of the Hamiltonian of Eq.~\eqref{eq: HS_gen_Hamiltonian}, i.e., $J^{\rm NN}_\alpha$, where
\begin{equation}
J^{\rm NN}_\alpha=\left(\frac{2\pi}{N}\right)^{\alpha}\left( \frac{1}{\left|\eta_p-\eta_{p+1}\right|} \right)^\alpha = \left(\frac{\left(\pi/N\right)}{\sin\left(\pi/N\right)}\right)^\alpha.
\label{eq: max_int_energy}
\end{equation}
Thus, at site $p$, the local field is small when $h_{p}/J^{\rm NN}_\alpha{\ll}1$, while it is large when $h_{p}/J^{\rm NN}_\alpha{\gg}1$. In the thermodynamic $N{\to}\infty$ limit, $J^{\rm NN}_\alpha{\to}1$ for all $\alpha$. 

\section{Results}
\label{sec: results}
We numerically analyzed the level statistics of the Hamiltonian of Eq.~\eqref{eq: z_h_dis_HS_gen_Hamiltonian} using eExact diagonalization (ED) in $U(1)$-symmetry resolved sectors characterized by total magnetization, $m$, that labels the $S_{\rm tot}^z{=}\sum_{1\leq p \leq N} S_j^z$ quantum number. For the clean Hamiltonian of Eq.~\eqref{eq: HS_gen_Hamiltonian}, we additionally use its translation invariance to resolve the pseudo momentum $k$, in units of $2\pi/N$ [the lattice spacing is set to unity]. For a given disorder realization, we compute the successive gap ratios, $\{\tilde{\mathfrak{r}}_n\}$ [see Eq.~\eqref{eq: ratio_consecutive_level_spacings}] and average them over the entire spectrum to obtain the mean gap ratio. This mean gap ratio is then further averaged over 1000 different disorder realizations to obtain the final level statistics parameter, denoted as $\langle \tilde{\mathfrak{r}}\rangle$. For convenience, the level statistics parameter for both the clean system (obtained from averaging $\{\tilde{\mathfrak{r}}_n\}$ over the spectrum) and the disordered system (obtained from averaging $\{\tilde{\mathfrak{r}}_n\}$ over the spectrum, as well as over disorder realizations) are denoted as $\langle \tilde{\mathfrak{r}}\rangle$. 

\begin{figure}[tbp!]
    \centering
    \includegraphics[width=1.0\linewidth]{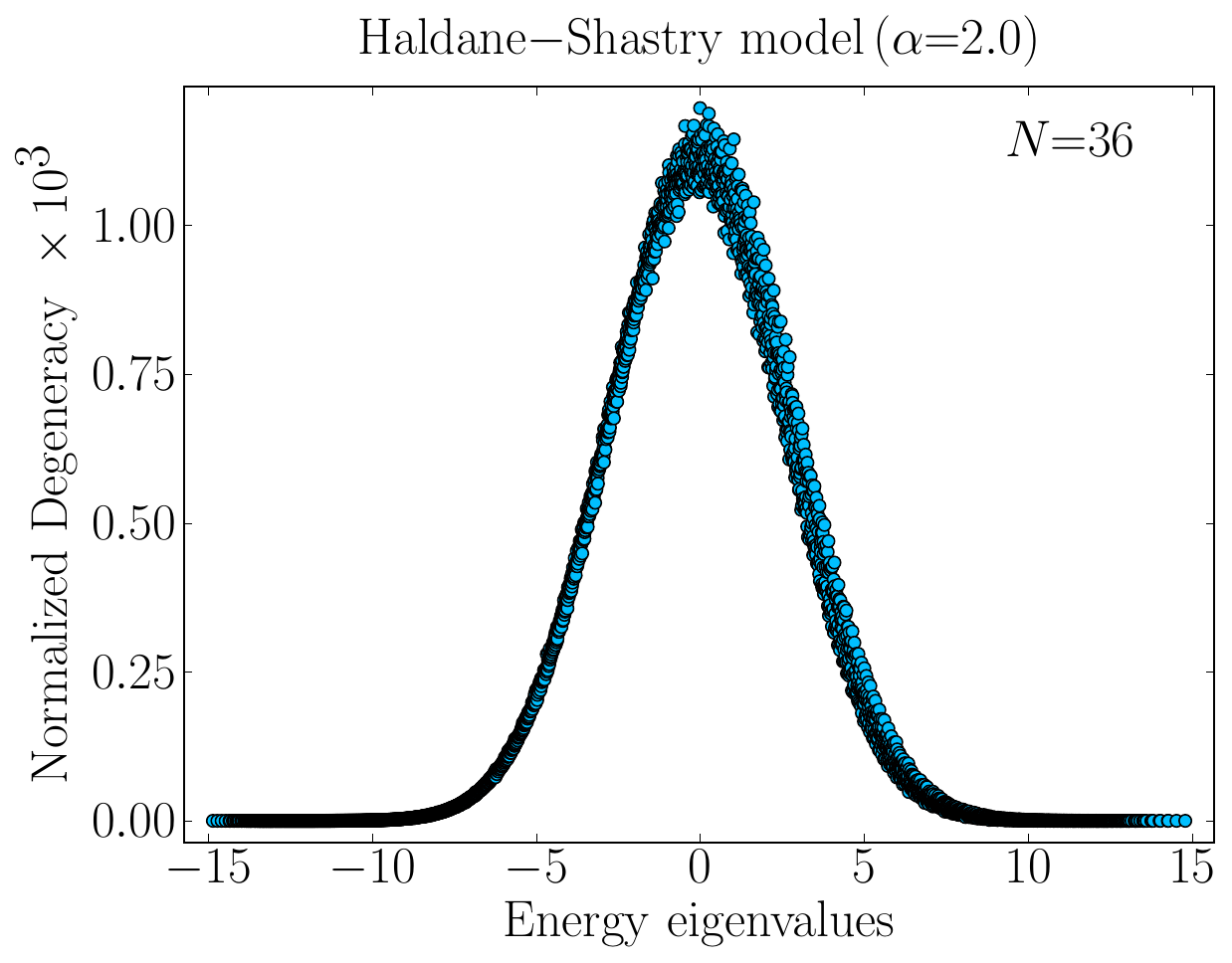}
    \caption{Normalized degeneracy of the full spectrum for $N{=}36$ spins of the $\alpha{=}2$ Haldane-Shastry model.}
    \label{fig: degeneracy_HS_model}
\end{figure}
\begin{figure*}[tbp!]
    \centering
    \includegraphics[width=1\textwidth]{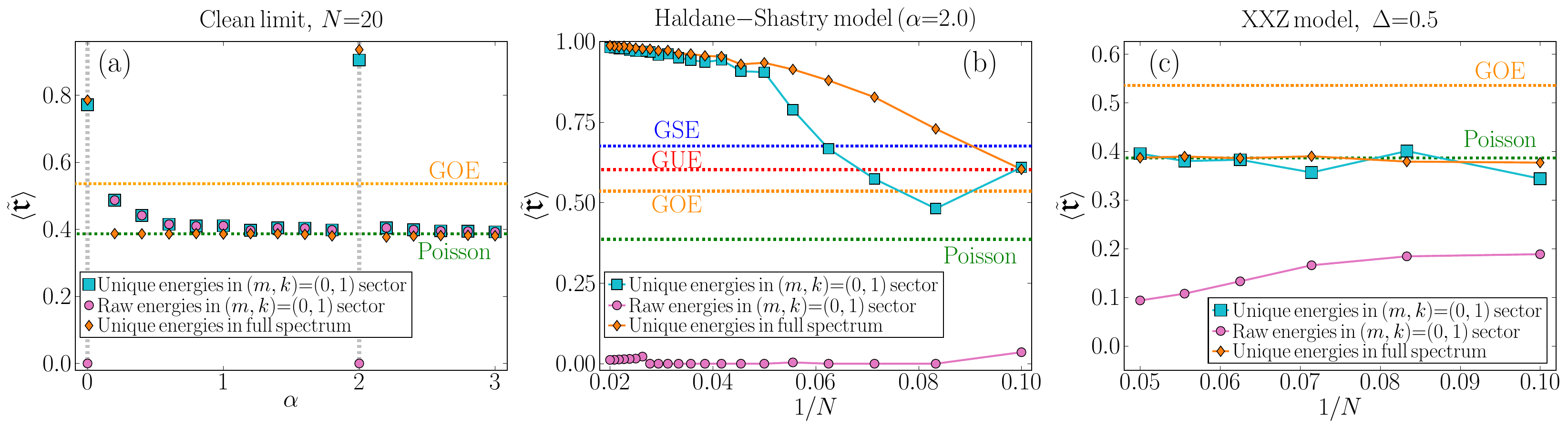}
    \caption{Analysis of the spectral-statistics before [raw energies] and after removing the degeneracies [unique energies]. (a) level statistics parameter $\langle \tilde{\mathfrak{r}}\rangle$ as a function of $\alpha$ for the clean system of Eq.~\eqref{eq: HS_gen_Hamiltonian} in the $(m,k){=}(0,1)$ symmetry sector and the full spectrum; (b,c) thermodynamic extrapolation of $\langle \tilde{\mathfrak{r}}\rangle$ as a function of $1/N$ (b) for the $\alpha{=}2$ Haldane-Shastry model of Eq. \eqref{eq: HS_gen_Hamiltonian} in the $(m,k){=}(0,1)$ sector and the full spectrum; (c) for comparison results of the XXZ Hamiltonian of Eq. \eqref{eq: XXZ Hamiltonian} at the anisotropic parameter $\Delta{=}\cos(\pi/3){=}0.5$ in the $(m,k){=}(0,1)$ sector and the full spectrum are shown.}
    \label{fig: statistics without degeneracies}
\end{figure*} 
The level statistics were computed for the interaction parameter $\alpha{\in}[0, 3]$, position disorder strengths $\delta{\in}[0,1]$ and random longitudinal field strengths $h/J^{\rm NN}_{\alpha}{\in}[0,4]$. In the limit $h/J^{\rm NN}_\alpha {\gg} 1$, the model becomes trivially or Anderson localized~\cite{Anderson58} as the magnetic field term $H_h$ dominates over the spin-spin interaction in this regime. For small values of $\delta$ and $h$, the translation symmetry of the clean model is approximately restored, which obfuscates the level statistics as in the case with exact symmetries, where the symmetries are not resolved. For these reasons, and factoring in the computing time available to us, we have chosen to restrict our analysis to the range of parameters given above. To assess finite-size effects, we consider system sizes $N{\in}\{ 12, 14, 16, 18, 20, 22\}$ for the clean system and $N{\in}\{10, 12, 14, 16\}$ for the disordered systems. As the $\alpha{=}2$ Haldane-Shastry model is exactly solvable, we calculated $\langle \tilde{\mathfrak{r}}\rangle$ for it for up to $N{=}50$ [see Fig. \ref{fig: statistics without degeneracies}(b)] using the method discussed in Ref. \cite{Greiter2007}. For convenience and to keep this article self-complete, we have outlined this method in the Appendix~\ref{app: HS_spectrum}.

\begin{figure*}[tbp!]
    \includegraphics[width=\textwidth]{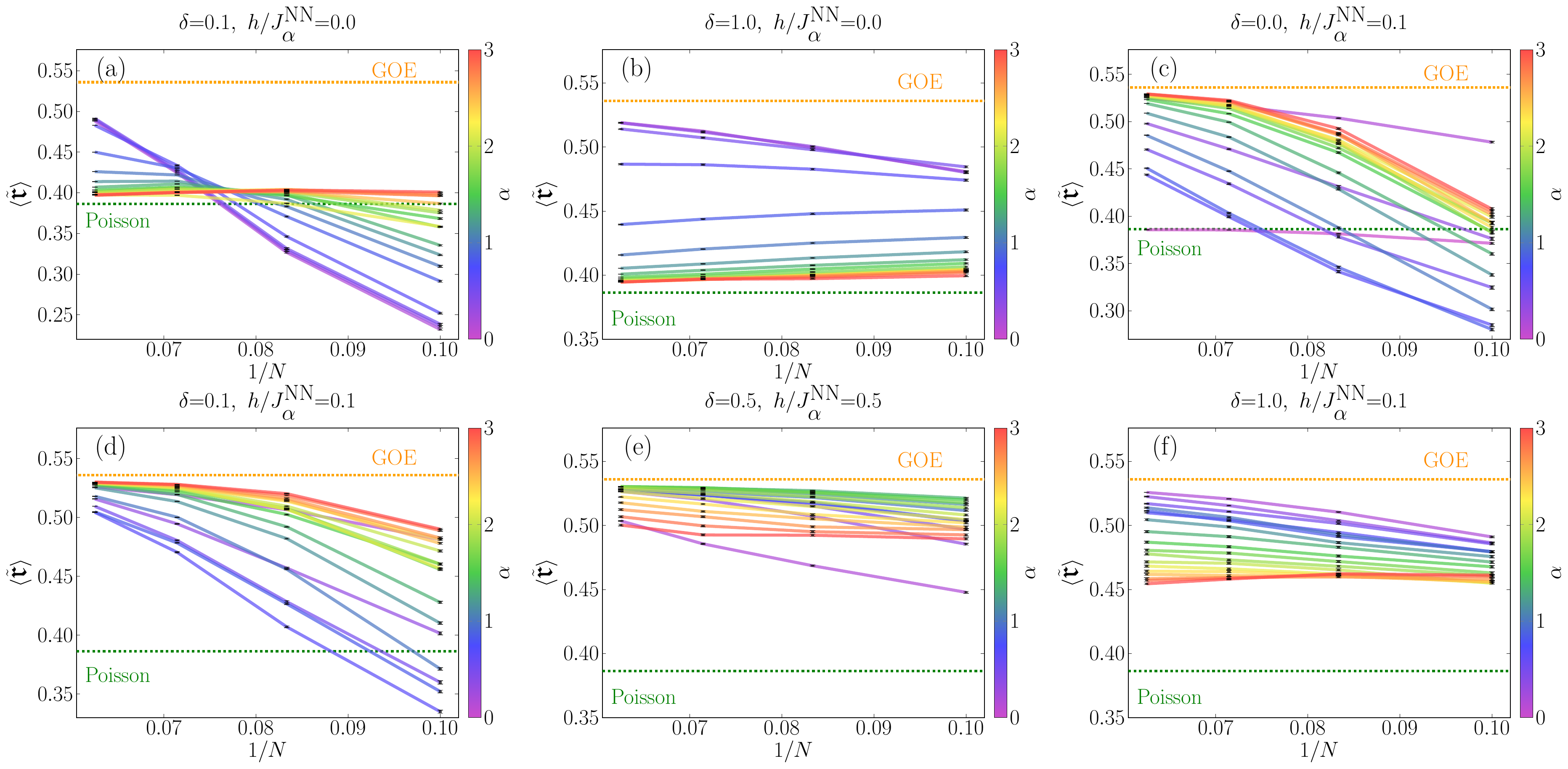} 
  \caption{\label{fig: z_and_h_dis_plots_extrapolation} Thermodynamic extrapolation of the level statistics parameter $\langle \tilde{\mathfrak{r}}\rangle$ of the Hamiltonian of Eq. \eqref{eq: z_h_dis_HS_gen_Hamiltonian} as a function of $1/N$ for interaction parameter $\alpha{\in}(0,3]$, disorder strength $\delta$ and magnetic field strength $h/J^{\rm NN}_\alpha$ in the zero magnetization sector. For each value of $\alpha$, $\delta$, and $h/J^{\rm NN}_\alpha$, the results are averaged over $1000$ independent disorder realizations. Results are shown for (a) $\delta{=}0.1$ and $h/J^{\rm NN}_\alpha{=}0.0$; (b) $\delta{=}1.0$ and $h/J^{\rm NN}_\alpha{=}0.0$; (c) $\delta{=}0.0$ and $h/J^{\rm NN}_\alpha{=}0.1$; (d) $\delta{=}0.1$ and $h/J^{\rm NN}_\alpha{=}0.1$; (e) $\delta{=}0.5$ and $h/J^{\rm NN}_\alpha{=}0.5$; (f) $\delta{=}1.0$ and $h/J^{\rm NN}_\alpha{=}0.1$. Considerable finite-size effects are seen for $\delta{=}0.1$ and/or $h/J^{\rm NN}_\alpha{=}0.1$, and these are particularly strong for $\alpha{<}1$ [see panels (a), (c) and (d)].}
\end{figure*}

The numerical results of $\langle \tilde{\mathfrak{r}}\rangle$ as a function of $\alpha$ for the largest system considered here, $N{=}22$, in the symmetry sector $(m,k){=}(0,1)$ for the clean system (as the largest $k{=}0$ sector commutes with parity, the second largest sector with $k{=}1$ is considered to avoid the additional parity symmetry) and $N{=}16$ in the $m{=}0$ sector for the disordered system are presented in Figs. \ref{fig: clean_model_plots}(a) and \ref{fig: z_h_dis_plots}. Our data is nicely converged, as evinced by the fact that the statistical standard errors in the mean are smaller than the marker size. The thermodynamic extrapolation of $\langle \tilde{\mathfrak{r}}\rangle$ is shown in Figs. \ref{fig: clean_model_plots}(b) and \ref{fig: z_and_h_dis_plots_extrapolation}. 

\subsection{Clean limit}
\label{ssec: clean_limit}
\begin{figure}[b]
    \centering
    \includegraphics[width=1\linewidth]{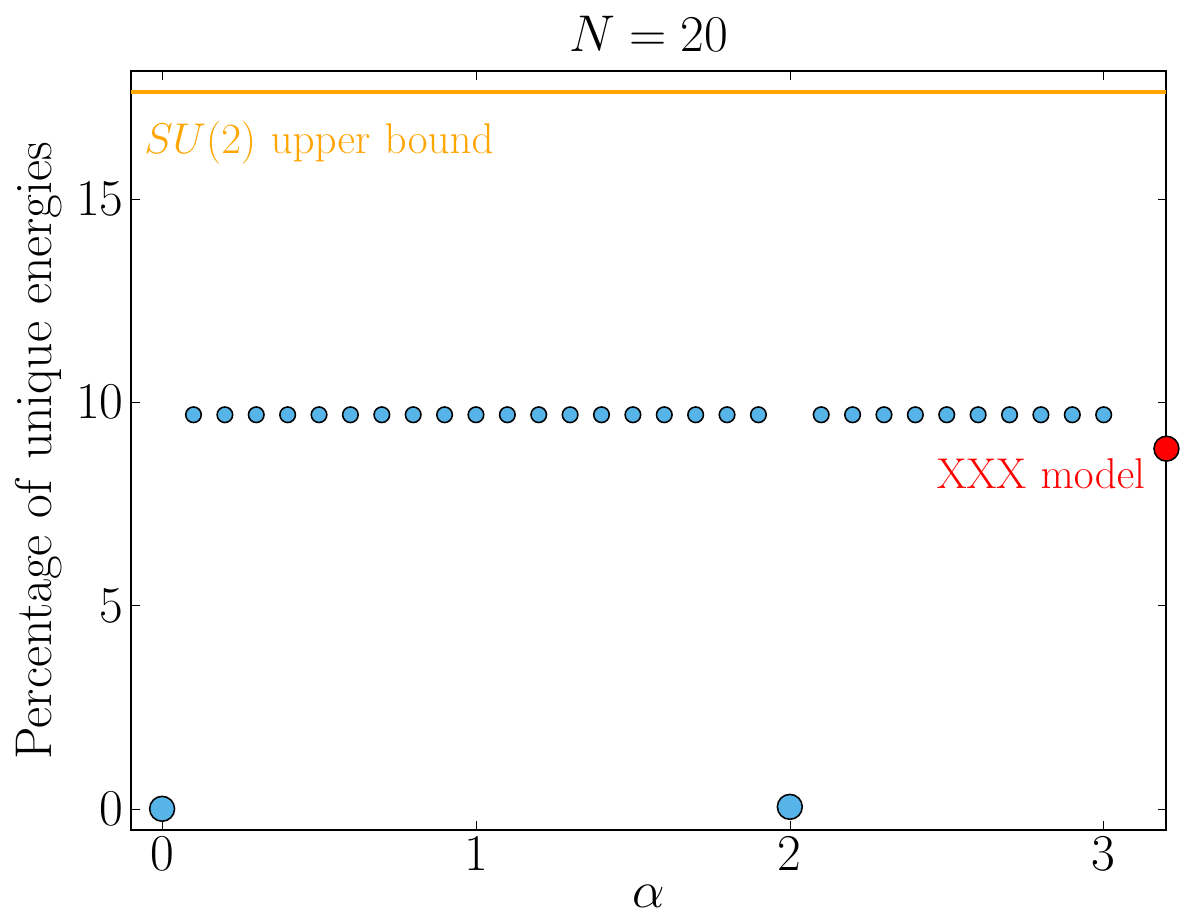}
    \caption{Percentage of unique energies as a function of $\alpha$ of the clean Hamiltonian of Eq.~\eqref{eq: HS_gen_Hamiltonian} for $N{=}20$. The red circle denotes the value for the XXX or Heisenberg model ($\alpha {\to} \infty$). }
    \label{fig: percentage_unique_energies_HS_gen}
\end{figure}

The clean Hamiltonian of Eq.~\eqref{eq: HS_gen_Hamiltonian} appears to show a transition from GOE to Poisson statistics as the interaction parameter $\alpha$ increases from $0$, except at the points $\alpha{=}0,2$, which show anomalous level statistics [see Fig.~\ref{fig: clean_model_plots}(a)]. For $\alpha{<}1$, the competing antiferromagnetic all-to-all interaction leads to frustration, which likely results in chaotic behavior as evidenced by the GOE statistics. When $\alpha{>}1$, the statistics are Poisson, even though for $\alpha{\gtrsim}1$, the nearby interactions are not substantially suppressed. This numerical observation, along with $\langle \tilde{\mathfrak{r}}\rangle$ extrapolating to values intermediate between Poisson and GOE for certain $\alpha$ in the thermodynamic limit, suggests that there is likely an unresolved symmetry, resolving which may require numerically filtering these matrices using spectral decimation~\cite{he2026spectral} or a bootstrap framework~\cite{bai2026bootstrappingsymmetriesquantummanybody}, that is worth considering in the future. Finite-size effects in the clean model appear to be negligible as $N$ increases for all the $\alpha$ we have considered, as evidenced by the smooth thermodynamic extrapolation shown in Fig.~\ref{fig: clean_model_plots}(b). Next, we discuss the anomalous behavior seen at the points $\alpha{=}0,2$.

For $\alpha{=}0$, the Hamiltonian of Eq.~\eqref{eq: HS_gen_Hamiltonian} becomes
\begin{equation}
    H= \frac{1}{\mathcal{N}}\sum_{1\leq p<q \leq N} \vec{S}_p \cdot \vec{S}_q 
      = \frac{1}{2\mathcal{N}} \left( \vec{S}_{\rm tot}^2 - NS(S+1) \right),
    \label{eq: HS_gen_Hamiltonian_alpha_0}
\end{equation}
where $\vec{S}_{\rm tot}{=}\sum_{p{=}1}^N \vec{S}_p$. Therefore, there are as many unique eigen-energies as there are possible values for the total spin $\vec{S}_{\rm tot}$. Viewing the Hamiltonian as a free single large spin with $\vec{S}_{\rm tot}$ quantum number $\mathbb{S}$, the corresponding energy eigenvalue is
\begin{equation}
\label{eq: HS_gen_energy_alpha_0}
    E(N,\mathbb{S}) = \frac{1}{2\mathcal{N}}\left( \mathbb{S}(\mathbb{S}+1) - \frac{3}{4}N \right),
\end{equation}
where $\mathbb{S}{=}N/2, N/2{-}1,{\cdots},0$ (Note that $N$ is even.). Thus, the energy-density scales linearly with $N$, implying $\gamma{=}1$, i.e., $\mathcal{N}{=}\mathcal{O}(N)$. The degeneracy of each level can be obtained by taking the difference between the number of states with successive $S^{z}_{\rm tot}$ values and factoring in the $2\mathbb{S}{+}1$ degeneracy for a given $\mathbb{S}$, which results in a net degeneracy of
\begin{equation}
    \label{eq: HS_gen_degeneracy_alpha_0}
    \mathcal{D}(N,\mathbb{S}) = (2\mathbb{S}+1)\left[\binom{N}{\frac{N}{2}-\mathbb{S}} - \binom{N}{\frac{N}{2}-\mathbb{S}-1}\right].
\end{equation}
Since the Hamiltonian is similar to that of a free (big) spin paramagnet, it is integrable. However, its level statistics do not match the Poisson one, owing to these large degeneracies given in Eq.~\eqref{eq: HS_gen_degeneracy_alpha_0}. 

The point at $\alpha{=}2$ is the HS model. The Yangian symmetry and quadratic dispersion [see Eq. \eqref{eq: spion_dispersion}] of the model results in significant degeneracies in its spectrum [see Fig. \ref{fig: clean_model_plots}(c)]. The number of unique energies grows polynomially~\cite{Finkel_Lopez_2015_Yangian_Fibonacci} despite the exponential growth of the Hilbert space. The upper bound for the number of unique states $\mathcal{N}_E$ is given by~\cite{Finkel_Lopez_2015_Yangian_Fibonacci} 
\begin{equation}
    \mathcal{N}_E=\frac{N}{12}(N^2+2)+1 \quad (N \mathrm{\,is\,even}).
    \label{eq: upper_bound_N_E_HS}
\end{equation}
The degeneracies force $\langle \tilde{\mathfrak{r}}\rangle {\approx} 0$, thereby rendering the statistics to be different from Poissonian even though the HS model is integrable. The density-of-states almost resembles a Gaussian distributed variable centered at zero energy~\cite{Finkel_2005} (see Fig. \ref{fig: degeneracy_HS_model}). However, this property is not exact as there is no particle-hole symmetry in the HS model, albeit there is an emergent one in a particular subspace as $N{\to}\infty$ due to vanishing $\mathcal{O}(1/N)$ corrections [see Appendix~\ref{app: emergent_PHS_HS}]. 

\subsection*{Level statistics with unique energies} 
The reason for the anomalous behavior in the level statistics of the clean model at $\alpha{=}0,2$ is due to the large degeneracies in its spectra. Many of these degeneracies arise from unresolved symmetries, as it is not always possible to resolve all symmetries numerically. Next, we compute the level statistics, keeping only the unique energies in the spectrum, to check whether it reveals anything about the model.

For $\alpha{=}0$, the successive energy differences are
\begin{equation}
    \label{eq: HS_energy_differences_alpha_0}
    d(N,\mathbb{S}) = E(\mathbb{S}{+}1)-E(\mathbb{S})=\mathbb{S}+1,
\end{equation}
where $\mathbb{S}{=}N/2{-}1,{\cdots},0$. Using the notation, as in the Introduction section, we have $d_{n}{=}n{+}1$, for $n{=}0,1,{\cdots}, N/2{-}1$. The ratio $\mathfrak{r}_{n}{=}\left(n{+}1\right)/n$, for $n{=}1,2,{\cdots},N/2{-}1$ and thus $\mathfrak{r}_n{>}1/\mathfrak{r}_{n}$. Therefore, the ratio of consecutive level spacings, $\tilde{\mathfrak{r}}_n{=}1/\mathfrak{r}_{n}$. The level statistics, which is the average value of $\tilde{\mathfrak{r}}_n$, is 
\begin{equation}
    \label{eq: HS_degeneracy_removed_level_statistics_alpha_0}
    \langle\tilde{\mathfrak{r}}\rangle_{\alpha=0}{=}\frac{1}{N/2{-}1}\sum_{n=1}^{N/2{-}1} \frac{n}{n+1}{=}\frac{N/2-H_{N/2}}{N/2-1},
\end{equation}
where $H_{n}{=}\sum_{t{=}1}^{n}1/t$ is the $n^{\rm th}$ Harmonic number. In the limit $N{\to}\infty$, we have $\langle\tilde{\mathfrak{r}}\rangle_{\alpha=0}{\to}1$, which signifies anomalous level statistics. Similarly, $\alpha{=}2$ also exhibits anomalous level statistics when considering unique energies in a symmetry sector [see Fig.~\ref{fig: statistics without degeneracies}(b)]. Interestingly, the values of $\langle\tilde{\mathfrak{r}}\rangle_{\alpha{=}0}$ and $\langle\tilde{\mathfrak{r}}\rangle_{\alpha{=}2}$ in the symmetry sector $(m,k){=}(0,1)$ are nearly the same as those of the full Hamiltonian when only the unique energies are considered [see Figs.~\ref{fig: statistics without degeneracies}(a-b)] and these values tend to unity as $N$ increases.

Motivated by these results, we calculate the level statistics keeping only the unique energies for other $\alpha$'s [see Fig.~\ref{fig: statistics without degeneracies}(a)]. The value of $\langle \tilde{\mathfrak{r}} \rangle$ in the $(m,k){=}(0,1)$ symmetry sector for $\alpha{>}1$ lies near the Poisson value, while for $\alpha{<}1$, it is in-between Poisson and GOE. This behavior is the same as that of the statistics with raw energies [see Fig.~\ref{fig: statistics without degeneracies}(a)]. As stated at the beginning of Sec.~\ref{ssec: clean_limit}, the in-between level statistics at $\alpha{<}1$ likely stems from the antiferromagnetic all-to-all frustrated interactions. At $\alpha=0$, this results in perfect degeneracies, as we demonstrated by rewriting the Hamiltonian as that of a single large spin. For the full Hamiltonian, after removing the degeneracies, the level statistics agree well with the Poisson for all $\alpha{\neq}0,2$. This can be understood by noting that the level statistics of the full Hamiltonian assemble energies from all symmetry sectors, effectively making the consecutive energies uncorrelated, leading to Poisson statistics. The degree of deviation between the statistics obtained with and without selecting only the unique energies can be quantified by the percentage of states that survive this filtering. For $\alpha{=}0$ and $2$, we have already seen that only a vanishing percentage of states remain post such filtering, whereas for $\alpha{\neq}0,2$, we find that a substantial number of states remain (see Fig.~\ref{fig: percentage_unique_energies_HS_gen}). For example, for $\alpha{=}3$, the percentage of unique energies is $9.7$ for $N{=}20$. The percentage of unique energies is less than that of the bound implied by the $SU(2)$ symmetry [the horizontal orange line in Fig.~\ref{fig: percentage_unique_energies_HS_gen}, which is at a value of $\binom{N}{N/2}/2^N{\times}100{=}17.6\%$] due to the presence of other symmetries like translation and parity. Aside from these, there appear to be no other degeneracies stemming from symmetries as evidenced by the fact that the statistics with raw and unique energies agree with each other in the $(m,k){=}(0,1)$ sector [see Fig.~\ref{fig: statistics without degeneracies}(a)].

As a benchmark for comparison, we analyze the integrable points of the following XXZ Hamiltonian of Eq.~\eqref{eq: XXZ Hamiltonian} for the anisotropic parameter $\Delta {=}(\tau {+} \tau^{-1}){/}2 {=} \cos(\theta)$, with $\tau$ being a root of unity, 
\begin{equation}
    H_{\text{XXZ}} = \sum_{1\leq p\leq N} \left( S_p^x S_{p+1}^x + S_p^y S_{p+1}^y + \Delta\, S_p^z S_{p+1}^z \right).
    \label{eq: XXZ Hamiltonian}
\end{equation}
At these integrable points, the model commutes with the $sl_2$ loop algebra, which makes it integrable~\cite{FabriciusMcCoy2001, Kashiwara2002, Deguchi2003}. For the point $\Delta{=}\cos(\pi/3){=}0.5$, the spectral statistics in a symmetry sector show similar behavior as the points at $\alpha{=}0,2$ in the clean limit of the generalized HS model. After removing the degeneracies, the level statistics are Poisson [see Fig. \ref{fig: statistics without degeneracies}(c)] and do not show anomalous behavior as seen at the $\alpha{=}0,2$ points of the generalized HS model.

These calculations indicate that the statistics post removal of the degeneracies also reveal the correct local correlation of the energy levels of the system, which implies that the $\alpha{=}0,2$ points of the generalized HS model inherently exhibit anomalous spectral statistics. This works especially well when states from different sectors overlap due to perfect degeneracies stemming from symmetries such as translation, and will not be as useful when sectors are well separated in energy.
\begin{figure}[tbp!]
    \centering
    \includegraphics[width=1\linewidth]{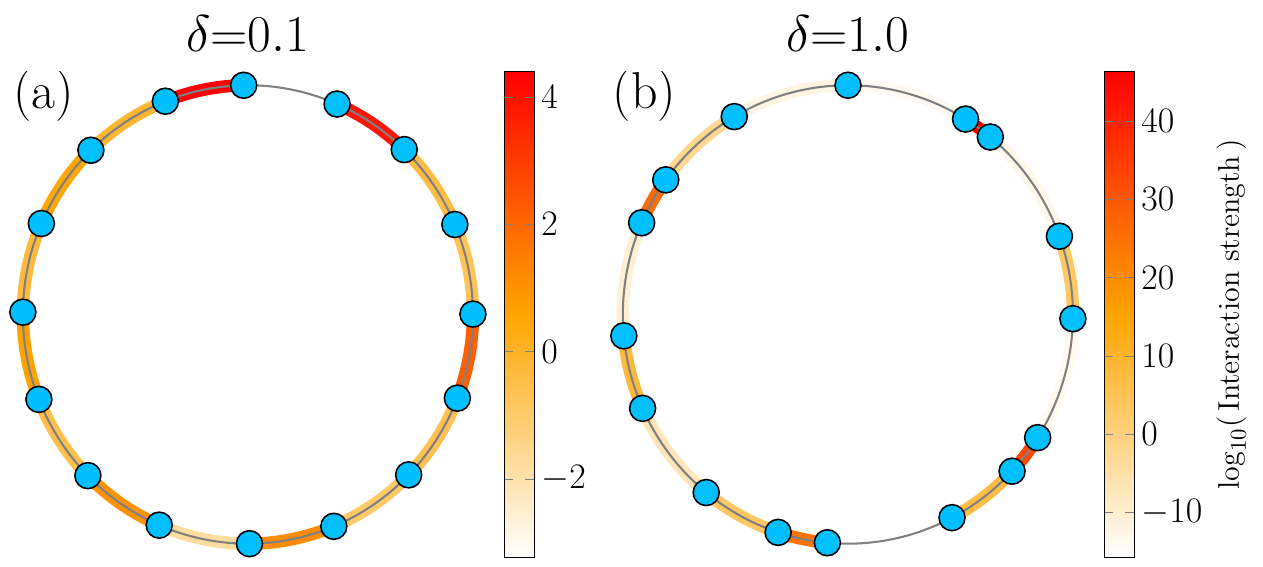}
    \caption{Interaction-strength profile of the position disordered Hamiltonian of Eq.~\eqref{eq: z_dis_HS_gen_Hamiltonian} at (a) $\delta {=}0.1$ (b) $\delta {=}1.0$; for interaction parameter $\alpha {=}100$ and $N{=}16$ spins. The red arc between the spins illustrates the interaction strength between them, with the opacity indicating the relative strength compared to the largest one. For $\alpha{=}100$, as the interaction is almost nearest-neighbor-like, even a small displacement away from the first neighbor makes the spins effectively disconnected, resulting in the formation of clusters of independent chains.}
    \label{fig: Lattice_alpha_100}
\end{figure}
\begin{figure*}[tbp!]
    \includegraphics[width=1\textwidth]{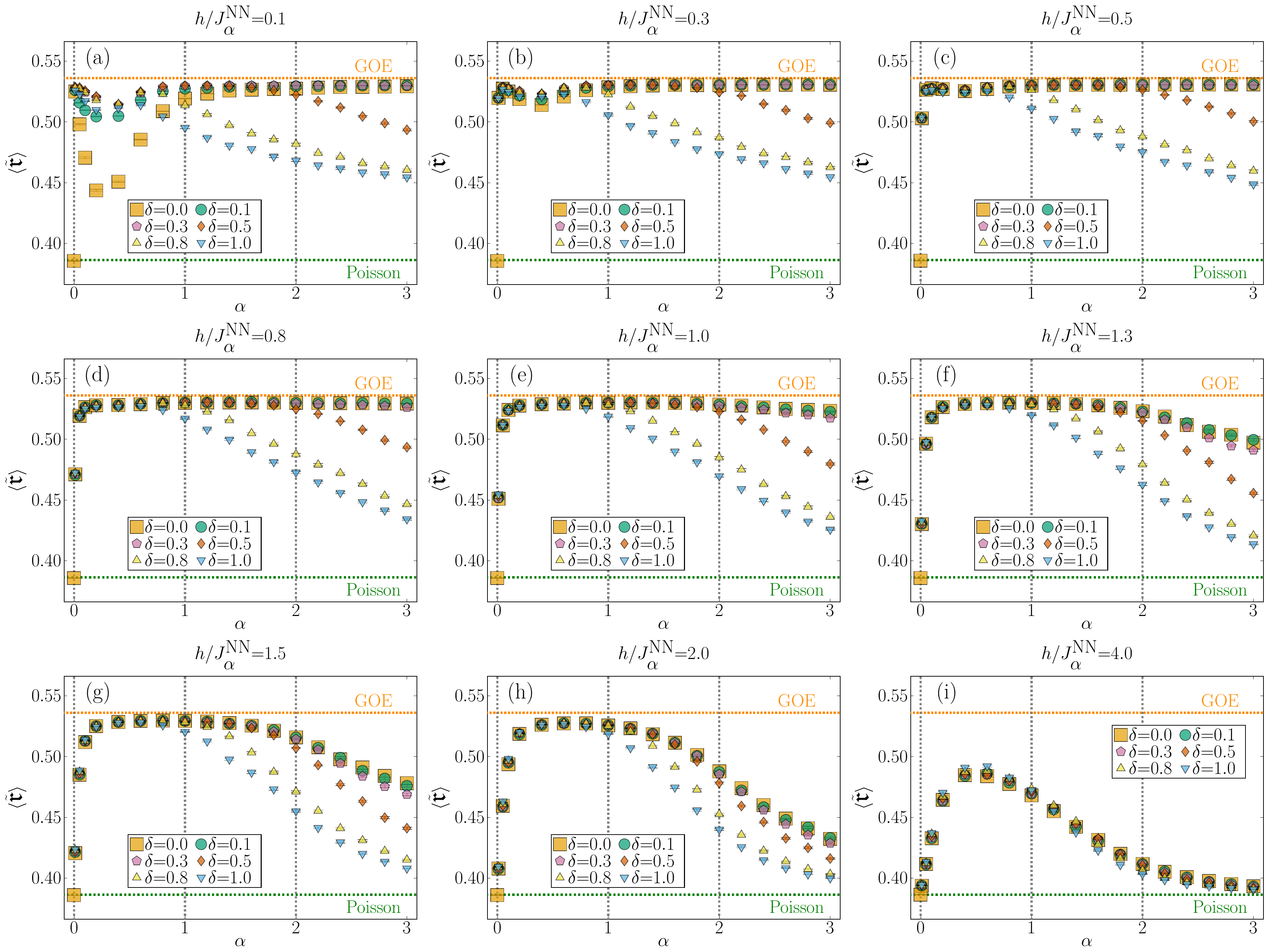}
  \caption{\label{fig: z_h_dis_plots} Level statistics parameter $\langle \tilde{\mathfrak{r}}\rangle$ of the Hamiltonian of Eq.~\eqref{eq: z_h_dis_HS_gen_Hamiltonian} as a function of the interaction parameter $\alpha$ for different values of position disorder strengths $\delta$ and magnetic field disorder strengths $h/J^{\rm NN}_\alpha$ in the zero magnetization sector. For each value of $\alpha$, $\delta$, and $h/J^{\rm NN}_\alpha$, the results are averaged over $1000$ independent disorder realizations. The gray vertical lines mark the points at $\alpha{=}0,1,2$. Results are shown for $N{=}16$ spins with (a) $h/J^{\rm NN}_\alpha{=}0.1$; (b) $h/J^{\rm NN}_\alpha{=}0.3$; (c) $h/J^{\rm NN}_\alpha{=}0.5$; (d) $h/J^{\rm NN}_\alpha{=}0.8$; (e) $h/J^{\rm NN}_\alpha{=}1.0$; (f) $h/J^{\rm NN}_\alpha{=}1.3$; (g) $h/J^{\rm NN}_\alpha{=}1.5$; (h) $h/J^{\rm NN}_\alpha{=}2.0$; (i) $h/J^{\rm NN}_\alpha{=}4.0$. The uniformity of $\langle \tilde{\mathfrak{r}}\rangle$ as a function of $N$, i.e., it being devoid of finite size effects [see Fig.~\ref{fig: z_and_h_dis_plots_extrapolation}], in the regime $h/J^{\rm NN}_\alpha{<}1$ and the drift towards the Poisson value in the regime $h/J^{\rm NN}_\alpha{\ge}1$ are evident.}
\end{figure*}
\begin{figure*}[tb!]
    \centering
    \includegraphics[width=1.0\linewidth]{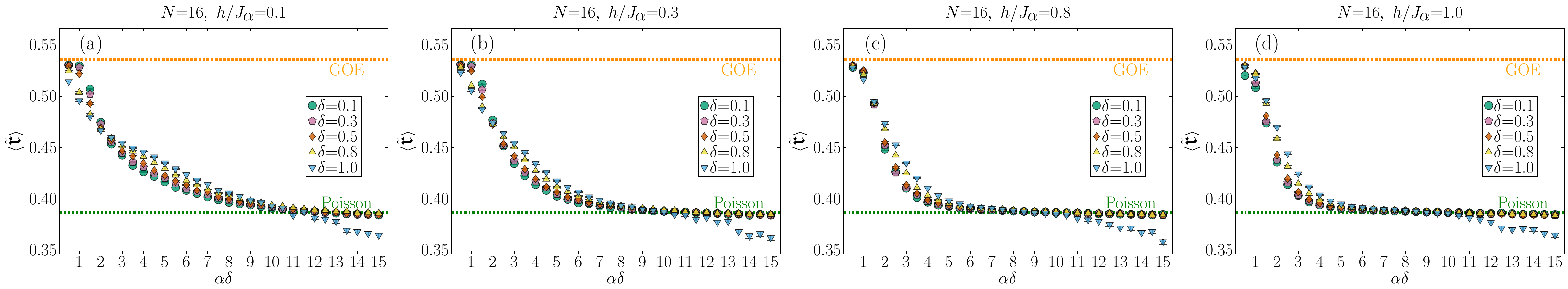}
    \caption{The average level statistics $\langle \tilde{\mathfrak{r}}\rangle$ of the Hamiltonian of Eq.~\eqref{eq: z_h_dis_HS_gen_Hamiltonian} as a function of the effective rescaled parameter $\alpha\delta$ in the zero magnetization sector for $N{=}16$ spins and different values of $\delta$ and  $h/J^{\rm NN}_\alpha$  (a) $h/J^{\rm NN}_\alpha{=}0.1$, (b) $h/J^{\rm NN}_\alpha{=}0.3$, (c) $h/J^{\rm NN}_\alpha{=}0.8$, and (d) $h/J^{\rm NN}_\alpha{=}1.0$.}
    \label{fig: rescaled_plot_alpha_delta_various_h}
\end{figure*}
\begin{figure*}[tb!]
    \centering
    \includegraphics[width=1.0\linewidth]{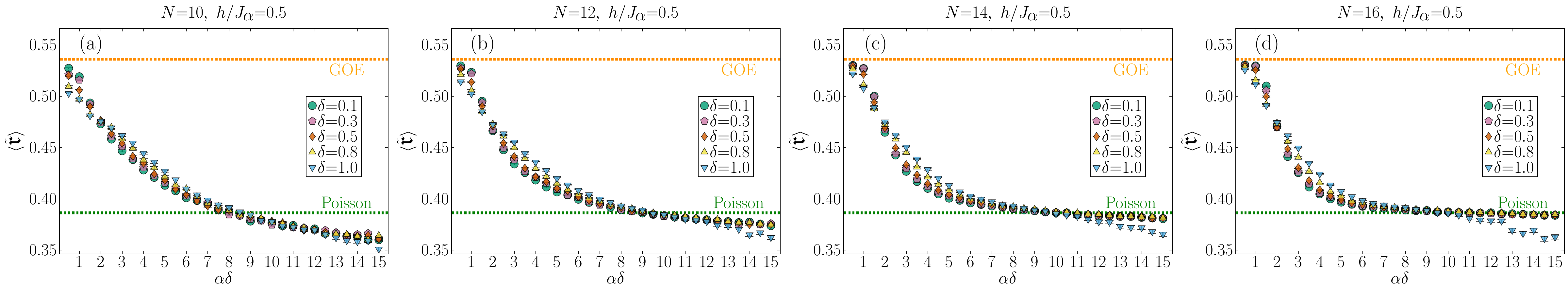}
    \caption{Average level statistics $\langle \tilde{\mathfrak{r}}\rangle$ of the Hamiltonian of Eq.~\eqref{eq: z_h_dis_HS_gen_Hamiltonian} as a function of the effective rescaled parameter $\alpha\delta$ for different values of $\delta$ and  $h/J^{\rm NN}_\alpha{=}0.5$ in the zero magnetization sector for (a) $N{=}10$, (b) $N{=}12$, (c) $N{=}14$, and (d) $N{=}16$ spins.}
    \label{fig: rescaled_plot_alpha_delta}
\end{figure*}

\subsection{Position disorder}
\label{ssec: position_disorder}

From this point onward, we only consider the raw energies without filtering out degeneracies in the zero magnetization sector. The position disordered Hamiltonian [see Eq.~\eqref{eq: z_dis_HS_gen_Hamiltonian}] appears to show a transition from GOE statistics to nearly Poisson as the interaction parameter $\alpha$ increases from $0$ [see Fig.~\ref{fig: clean_model_plots}(a)]. The disorder-averaged gap ratio $\langle \tilde{\mathfrak{r}}\rangle$ follows the same trend as the clean one [see Fig.~\ref{fig: clean_model_plots}(a)], with the results being largely insensitive to the disorder strength, once it is non-zero. For $1{\leq}\alpha{\leq}3$, $\langle \tilde{\mathfrak{r}}\rangle$ saturates at a value slightly higher than Poisson [see Figs.~\ref{fig: z_and_h_dis_plots_extrapolation}(a) and~\ref{fig: z_and_h_dis_plots_extrapolation}(b)]. This behavior is consistent with observations reported in Ref.~\cite{Shriya_Pai_2017} for $\alpha {=} 2$, further supporting the incompatibility of area-law scaling of entanglement entropy of eigenstates in the presence of $SU(2)$ symmetry. This does not apply in the clean system, where the integrability drives the statistics to be Poisson (except for the anomalous statistics at $\alpha{=}2$), as we have seen in the previous section. When $\alpha{\gg}1$, even a small displacement away from the first neighbor can cause the interaction strength to be negligible, thus effectively disconnecting the chain into smaller sub-chains (see Fig.~\ref{fig: Lattice_alpha_100}). The larger the $\delta$, the smaller the value of $\alpha$ at which this fragmentation of the chain happens. As the distance between the nearest-neighbor spins due to the position disorder becomes less than the corresponding nearest-neighbor distance in the clean limit, the interaction strength rapidly increases, making numerical computations unstable. 

\subsection{Magnetic field disorder}
\label{ssec: magnetic_field_disorder}
The magnetic field disordered Hamiltonian appears to exhibit GOE statistics for all $\alpha$ when $h/J^{\rm NN}_\alpha{<}1$, except at $\alpha{=}0$ [see Figs. \ref{fig: z_h_dis_plots}(b-d)]. This behavior indicates that the field strength is not strong enough to induce localization; therefore, the system remains in the ergodic phase. As $h/J^{\rm NN}_\alpha{\gtrsim}1$, the magnetic field term starts to dominate and the spectral statistics drift toward the Poisson [see Figs. \ref{fig: z_h_dis_plots}(e)-\ref{fig: z_h_dis_plots}(i)]. For $\alpha{=}0$, we have shown previously that the spectrum is given by a few discrete energy values, which are well-separated from each other [see Eq.~\eqref{eq: HS_energy_differences_alpha_0}] and each of which has a large degeneracy that at least scales as $N$ [see Eq.~\eqref{eq: HS_gen_degeneracy_alpha_0}]. In each of these levels, the effect of $H_h$ is to split the degeneracy by shifting the energy levels randomly. 
As a result, the successive energy level spacings of the $\alpha{=}0$ model with the random longitudinal field exhibit the same statistical properties as those of $H_h$ alone, which is Poisson. 

Although the magnetic field breaks the conventional time reversal symmetry generated by $T_0$, the spectral statistics remain of the GOE type. This is because of the presence of an unconventional time reversal symmetry, generated by $e^{i\pi S_x} T_0$~\cite{Haake2018Quantum, Avishai2002}. The unitary operator $e^{i\pi S_x}$ reverses the signs of $S_y$ and $S_z$; therefore, when combined with $T_0$, it effectively reverses only the sign of $S_x$, leaving $H_h$ invariant. This unconventional time reversal symmetry ensures that the Hamiltonian can be represented as a real matrix in an appropriate basis, thereby leading to GOE statistics. 

\subsection{Both position and magnetic field disorders}
\label{ssec: position_magnetic_field_disorders}

The Hamiltonian with both types of disorder [see Eq.~\eqref{eq: z_h_dis_HS_gen_Hamiltonian}] appears to exhibit a transition from GOE to Poisson statistics in the regime $h/J^{\rm NN}_\alpha{<}1$ as $\alpha$ increases, suggesting a crossover from a delocalized to a localized phase. The nature of the transition depends mainly on $\delta$, while remaining largely independent of $h$ [see Figs.~\ref{fig: z_h_dis_plots}(b-d)]. In the presence of only position disorder, the transition behavior remains similar for different $\delta$. However, introducing a random magnetic field breaks this similarity. Based on the numerical results presented in Fig.~\ref{fig: z_h_dis_plots}, the point $\alpha$ at which $\langle \tilde{\mathfrak{r}} \rangle$ deviates from the GOE value is inversely proportional to $\delta$. This suggests that the transition could be characterized in terms of the combined single-parameter $\alpha \delta$. In Figs.~\ref{fig: rescaled_plot_alpha_delta_various_h} and~\ref{fig: rescaled_plot_alpha_delta} , we show $\langle \tilde{\mathfrak{r}} \rangle$ as a function of $\alpha\delta$ and find an approximate scaling collapse. The fluctuations for $\delta {=}1$ as the system approaches the Poisson limit stem from the numerical instability and clustering issues explained in Sec.~\ref{ssec: position_disorder}.  As $h/J^{\rm NN}_\alpha$ increases beyond unity, the $\langle \tilde{\mathfrak{r}} \rangle$ slowly drifts towards the Poisson value [see Figs. \ref{fig: z_h_dis_plots}(e-i)] as $H_h$ begins to dominate the Hamiltonian in Eq. \eqref{eq: z_h_dis_HS_gen_Hamiltonian} resulting in single-particle Anderson localization. Fig. \ref{fig: z_h_dis_plots}(i) clearly shows that for large $h/J^{\rm NN}_\alpha$ we do not expect any scaling in the variable $\alpha\delta$ as the numerical results show no significant dependence on $\delta$. These claims are further validated through the inverse participation ratio (IPR) analysis (see Appendix~\ref{app: ipr}).

\section{Conclusions}
\label{sec: conclusions}
In this work, we studied the level statistics of a variant of the spin $1/2$ Haldane-Shastry Model [Eq.~\eqref{eq: HS_gen_Hamiltonian}] in the presence of tunable long-range interaction, position disorder, random magnetic fields, and their combined effects.

In the absence of disorder, the Hamiltonian appears to exhibit a clear distinction between the long-range ($\alpha {<} 1$) and short-range ($\alpha {>}1$) regimes in its spectral properties, characterized by a transition from GOE to Poisson statistics, with $\alpha {=}0,2$ being outliers. Interestingly, even near the onset of the short-range regime, where interactions are not strictly limited to nearest neighbors, the spectral statistics follow Poisson behavior. This observation suggests that there could be an unresolved symmetry in this parameter regime, an investigation of which we leave to future work. The anomalous behavior at $\alpha {=}0,2$ is due to the enormous degeneracies in the spectrum. At $\alpha{=}0,2$, the level statistics after removing the degeneracies from the spectrum also show anomalous behavior, different from other integrable models with degeneracies, indicating that these do not fit into the RMT picture.

When position disorder alone is introduced, the system qualitatively displays the same trend as in the clean case across all considered disorder strengths. However, instead of approaching Poisson statistics, the level spacing ratio saturates at a value slightly higher than the Poisson limit. This behavior persists throughout the short-range regime ($\alpha {>} 1$), and therefore no localization transition is observed. These findings are consistent with previous results reported in Ref.~\cite{Shriya_Pai_2017} for the position-disordered $\alpha{=}2$ Haldane-Shastry Model and further support the incompatibility of the $SU(2)$ symmetry with MBL and area law entanglement scaling of eigenstates.

As the $SU(2)$ symmetry of the Hamiltonian in Eq. \eqref{eq: z_dis_HS_gen_Hamiltonian} is broken down to $U(1)$ by a random magnetic field, the system appears to exhibit GOE statistics for $h/J^{\rm NN}_\alpha{<}1$ for all $\alpha$, indicating an ergodic phase. As $h/J^{\rm NN}_\alpha{\gtrsim}1$, the spectral statistics transition from GOE to Poisson, with the larger $\alpha$ exhibiting a faster drift toward Poisson statistics.

In the presence of both position and magnetic field disorder, for $h/J^{\rm NN}_\alpha{<}1$, the spectral statistics appear to exhibit a clear crossover from GOE to Poisson statistics, signaling the onset of MBL. The point $\alpha$ at which $\langle \tilde{\mathfrak{r}} \rangle$ deviates from GOE statistics is inversely proportional to $\delta$, while remaining largely independent of $h$. This highlights the important role of position disorder in driving localization once the $SU(2)$ symmetry is explicitly broken down to $U(1)$. This inverse relationship between $\alpha$ and $\delta$ suggested a new control parameter, $\alpha \delta$, under which data for $\langle \tilde{\mathfrak{r}} \rangle$ for different $\delta$ and $\alpha$ could collapse onto a single universal curve. We do find such an approximate scaling collapse, suggesting that the crossover behavior can be captured by a single effective parameter, $\alpha \delta$. In particular, even at the maximum strength of the position disorder, the spectral statistics remain strictly GOE in the long-range regime ($\alpha{<}1$). As $h/J^{\rm NN}_\alpha{\gtrsim}1$, the value of $\langle \tilde{\mathfrak{r}} \rangle$ slowly drifts towards the Poisson value, as is expected from single-particle Anderson localization for strong random magnetic fields.

Overall, our results demonstrate that in the parameter regime where the spin-spin interaction dominates over the magnetic field ($h/J^{\rm NN}_\alpha{<}1$), neither position disorder nor magnetic field disorder alone is sufficient to induce Poisson statistics in this long-range interacting system. Instead, Poisson statistics arise only from their combined presence, suggesting the emergence of a MBL.

Finally, we note that our analysis is based solely on spectral statistics. Although these provide a robust indicator of ergodicity breaking, they may not capture other key properties of MBL, such as entanglement scaling, dynamical, and transport behavior. Exploring complementary diagnostics, such as the spectral form factor and entanglement entropy scaling, would be useful to fully characterize the localized regime in long-range interacting systems and to assess both the stability of the observed transition and the nature of the crossover region. In the future, it would be interesting to do a similar analysis of the Inozemtsev model \cite{Inozemtsev1990JSP}, which interpolates continuously between the Haldane–Shastry and XXX models while always remaining integrable.

\begin{acknowledgments}
We acknowledge valuable discussions with N S Srivatsa. The work was made possible by financial support from the Science and Engineering Research Board (SERB) of the Department of Science and Technology (DST) via the Mathematical Research Impact Centric Support (MATRICS) Grant No. MTR/2023/000002 and the Anusandhan National Research Foundation (ANRF) via the Advanced Research Grant No. ANRF/ARG/2025/000562/PS. Computational portions of this work were undertaken on the Kamet supercomputer, which is maintained and supported by the Institute of Mathematical Sciences' High Performance Computing Center. The results for the spin-$1$ system were obtained using QuSpin~\cite{Quspin1, Quspin2}.
\end{acknowledgments} 

\section*{Data availability} 
The data that support the findings of this article are openly available~\cite{haldane_shastry_repo}.

\appendix
\begin{figure}[tb]
    \centering
    \includegraphics[width=1.0\linewidth]{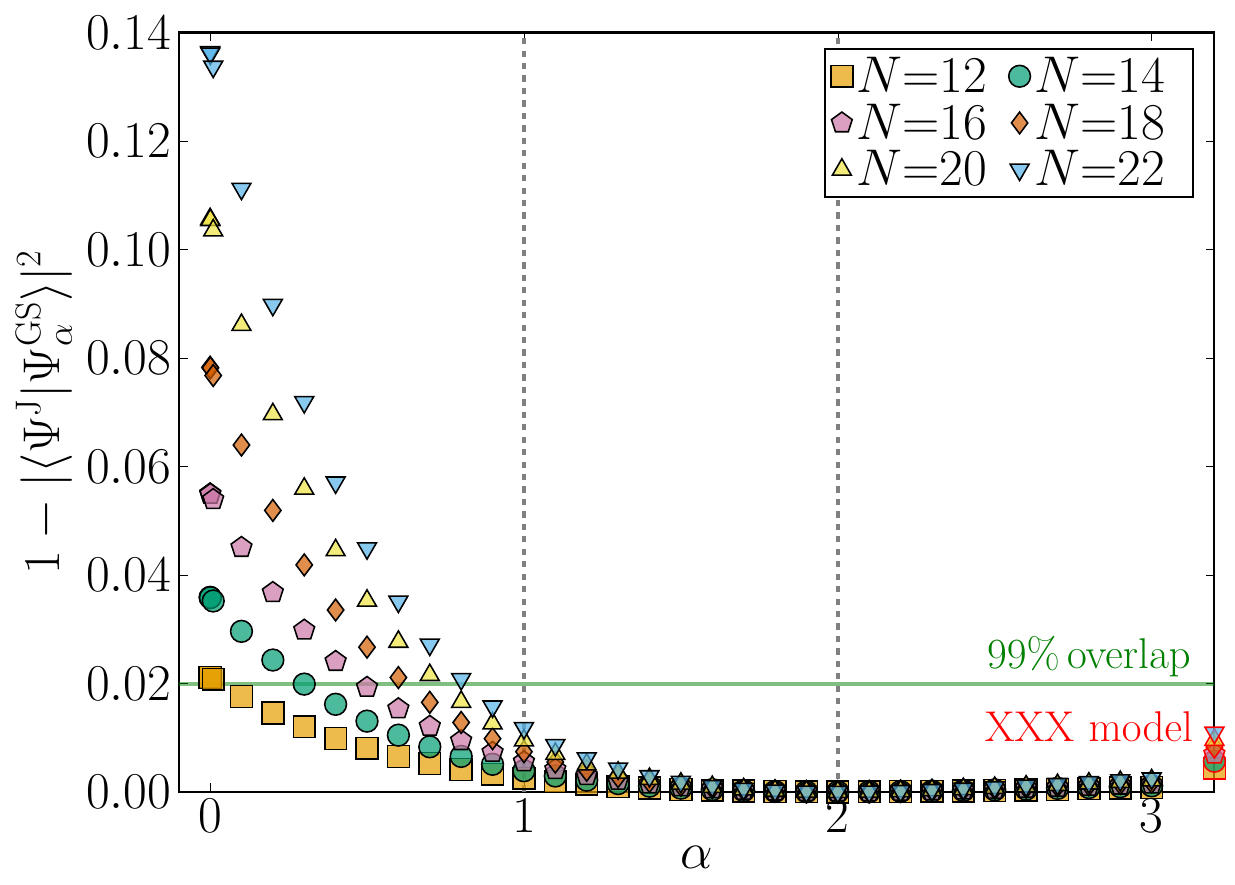}
    \caption{Squared overlap deviation from unity, i.e., $1-|\langle{\Psi^{\mathrm{J}}}|{\Psi_{\alpha}^{\mathrm{GS}}}\rangle|^{2}$, as a function of $\alpha$ for the clean system for various systems, where $|{\Psi_{\alpha}^{\mathrm{GS}}}\rangle$ is the numerically evaluated ground state at $\alpha$ and $|{\Psi^{\mathrm{J}}}\rangle$ is the ground state at $\alpha{=}2$ [defined in Eqs.~\eqref{eq: HS_Groundstate} and~\eqref{HS_wavefunction}]. The non-interacting ground state at $\alpha{=}0$ has low overlap with $|{\Psi^{\mathrm{J}}}\rangle$, and has been omitted for clarity. The gray vertical lines mark the points at $\alpha{=}1,2$, and the markers at the right end with red borders denote the results for the $\alpha{\to}\infty$ XXX Heisenberg model.
}
    \label{fig:Jastrow clean HS gen}
\end{figure}
\begin{figure*}[tb]
    \centering
    \includegraphics[width=1\linewidth]{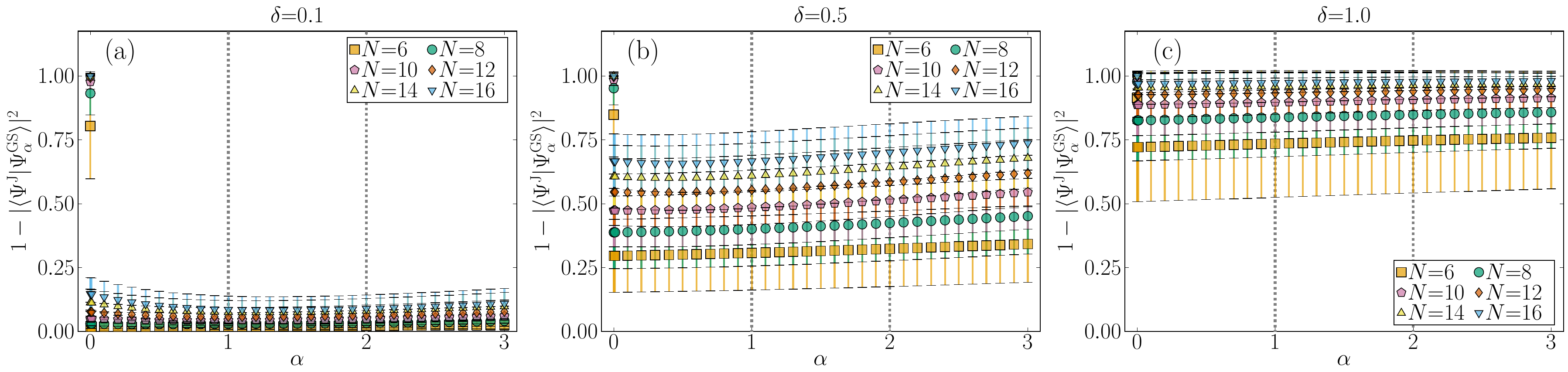}
    \caption{Mean and standard deviation of the overlap deviation, $1{-}|\langle{{\Psi^{\mathrm{J}}}}|{\Psi_{\alpha}^{\mathrm{GS}}}\rangle|^{2}$, where $|{\Psi^{\mathrm{J}}}\rangle$ is the state defined in Eqs.~\eqref{eq: HS_Groundstate} and~\eqref{HS_wavefunction} with $\{\eta\}$ being the (disordered) positions on the unit circle, and $|{\Psi_{\alpha}^{\mathrm{GS}}}\rangle$ is the numerically evaluated ground state for the position disordered system at $\alpha$, over $1000$ independent disorder realizations as a function of $\alpha$ for various system sizes and disorder strengths $\delta$, (a) $\delta{=}0.1$, (b) $\delta{=}0.5$, (c) $\delta{=}1.0$. The gray vertical lines mark the points at $\alpha{=}1,2$. The error bars show the standard deviation in the distribution of squared overlap values.
    }
    \label{fig:Jastrow overlap disordered HS gen}
\end{figure*}

\section{Relevance of the Jastrow wavefunction in the Haldane-Shastry model with $1/r^\alpha$ interaction}
\label{app: Jastrow}

\begin{figure*}[t]
    \centering
    \includegraphics[width=1\linewidth]{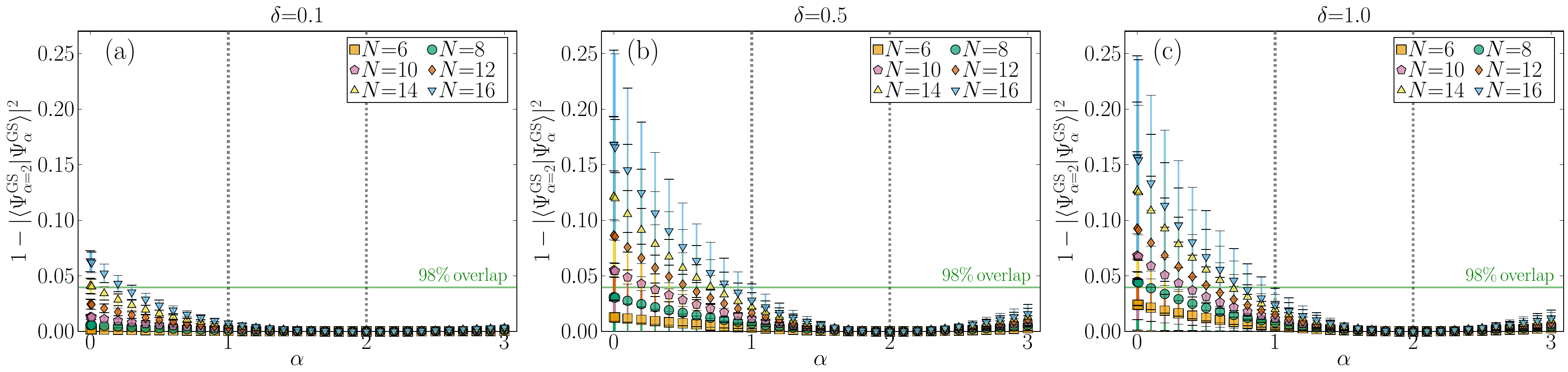}
    \caption{Mean and standard deviation of the overlap deviation, $1{-}|\langle{\Psi_{\alpha=2}^{\mathrm{GS}}}|{\Psi_{\alpha}^{\mathrm{GS}}}\rangle|^{2}$, where $|{\Psi_{\alpha}^{\mathrm{GS}}}\rangle$ is the numerically evaluated ground state for the position disordered system at $\alpha$, over 1000 independent disorder realizations as a function of $\alpha$ for the position disordered system for various system sizes and disorder strengths $\delta$, (a) $\delta{=}0.1$, (b) $\delta{=}0.5$, (c) $\delta{=}1.0$.}
    \label{fig: HS overlap disordered HS gen}
\end{figure*}
\begin{figure*}[b]
    \centering
    \includegraphics[width=1\linewidth]{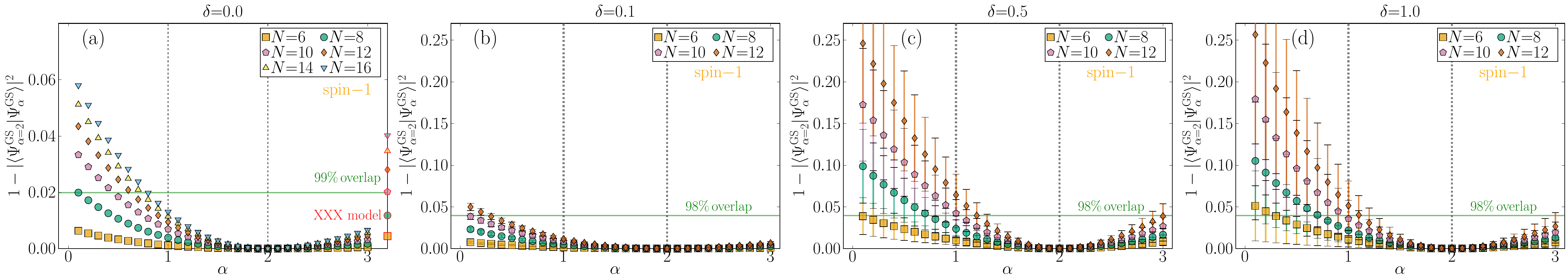}
    \caption{Mean and standard deviation of the overlap deviation, $1{-}|\langle{\Psi_{\alpha=2}^{\mathrm{GS}}}|{\Psi_{\alpha}^{\mathrm{GS}}}\rangle|^{2}$, where $|{\Psi_{\alpha}^{\mathrm{GS}}}\rangle$ is the numerically evaluated ground state for the clean and position disordered system [see Hamiltonian defined in Eq.~\eqref{eq: z_dis_HS_gen_Hamiltonian}] for spin $S{=}1$ at interaction parameter $\alpha$, over 1000 independent disorder realizations as a function of $\alpha$ for the position disordered system for various system sizes and disorder strengths $\delta$, (a) $\delta{=}0.0$ [clean system], (b) $\delta{=}0.1$, (c) $\delta{=}0.5$, and (d) $\delta{=}1.0$.}
    \label{fig: HS overlap spin 1}
\end{figure*}
The $\alpha{=}2$ HS model~\cite{Haldane88, Shastry88} admits an exact ground state that takes the following Jastrow form
\begin{equation}
    |{\Psi^{\mathrm{J}}}\rangle = \sum_{\{\eta_1, \ldots, \eta_M\}} \psi^{\mathrm{J}}(\eta_1, \ldots, \eta_M) S_{\eta_1}^+ \cdots S_{\eta_M}^+ \,\, |{\underbrace{\downarrow \downarrow \cdots \downarrow \downarrow}_{\text{all } N \text{ spins } \downarrow} }\rangle.
    \label{eq: HS_Groundstate}
\end{equation}
The sum in Eq.~\eqref{eq: HS_Groundstate} is taken over all possible configurations of $M{=}N/2\,\,\uparrow $-spin coordinates $\eta_p$ on the unit circle, and
\begin{equation}
    \psi^{\mathrm{J}}(\eta_1, \ldots, \eta_M)=\prod^M_{p<q} (\eta_p\,-\,\eta_q)^2\,\prod^M_{p=1}\eta_p.
    \label{HS_wavefunction}
\end{equation}
Interestingly, the ground state of the clean Hamiltonian of Eq.~\eqref{eq: HS_gen_Hamiltonian} preserves this structure for $\alpha{>}1$ as evidenced by overlaps that are upwards of $99\%$ with the wavefunction of Eq.~\eqref{eq: HS_Groundstate} in this short-range regime (see Fig.~\ref{fig:Jastrow clean HS gen}) for all the systems considered. For the disordered Hamiltonian of Eq.~\eqref{eq: z_dis_HS_gen_Hamiltonian}, as the disorder strength increases, the ground state deviates from the Jastrow form (see Fig.~\ref{fig:Jastrow overlap disordered HS gen}). We note that for $\alpha{=}2$, addition of another term in the Hamiltonian of Eq.~\eqref{eq: HS_Groundstate} ensures that the Jastrow form is the exact ground state so long as the (disordered) positions $\{\eta\}$ reside on the unit circle~\cite{Nielsen_2011}. Surprisingly, in the short-range regime of $\alpha{>}1$, the overlap between its exact ground state and that of the ground state at $\alpha{=}2$ continues to remain upwards of $98\%$ [see Fig.~\ref{fig: HS overlap disordered HS gen}] for the same disorder realization for all the system sizes considered. These results suggest that the low-energy physics in the entire short-range regime of $\alpha{>}1$ likely resides in the same universality class. 

This is similar in spirit to the idea that the Jastrow-form wavefunction, proposed by Laughlin~\cite{Laughlin83}, that penalizes the short-range repulsion~\cite{Haldane83, Trugman85}, lies in the same universality class as the exact Coulomb ground state of certain fractional quantum Hall fluids. Mapping spin-1/2 operators to hard-core bosons, where up spins correspond to occupied bosons and down spins to empty sites, singlet states correspond to a half-filled bosonic lattice, naturally providing a platform to realize the bosonic Laughlin state~\cite{Laughlin83} at filling fraction 1/2. We note that the Jastrow form also provides a highly accurate variational state for other spin systems, such as the resonating valence bond state for the Heisenberg antiferromagnet on a triangular lattice~\cite{Kalmeyer87, Kalmeyer89}, square lattice~\cite{Zou89}, the valence bond solid in the Affleck-Kennedy-Lieb-Tasaki model~\cite{Arovas88}, etc.

\begin{figure*}[tbp]
    \centering
    \includegraphics[width=1\linewidth]{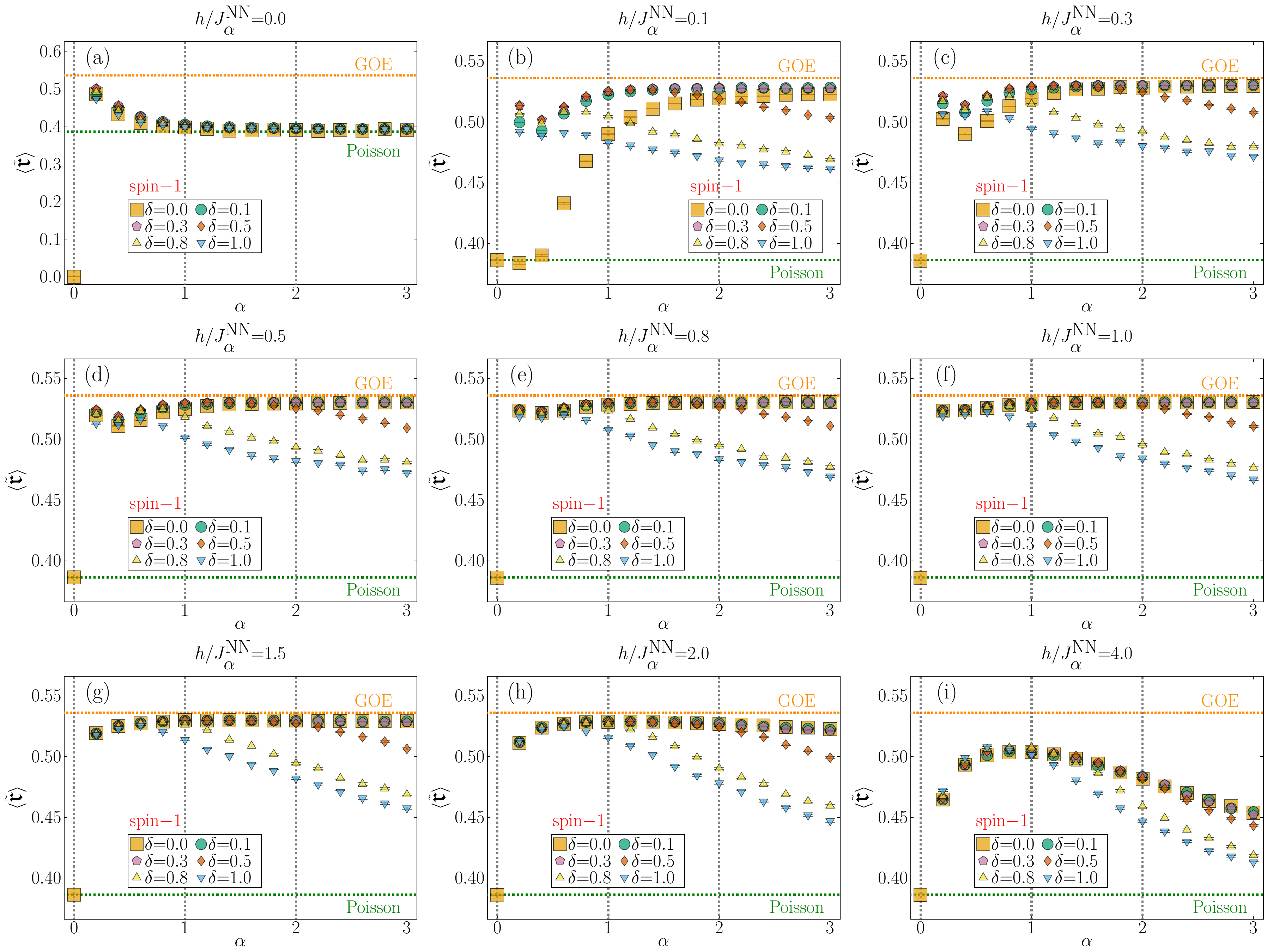}
    \caption{Level statistics parameter $\langle \tilde{\mathfrak{r}}\rangle$ of the Hamiltonian of Eq.~\eqref{eq: z_h_dis_HS_gen_Hamiltonian} for spin $S{=1}$ system as a function of the interaction parameter $\alpha$ for different values of position disorder strengths $\delta$ and magnetic field disorder strengths $h/J^{\rm NN}_\alpha$ in the zero magnetization sector and additionally in pseudo-momentum $k{=}1$ sector for the clean system. For each value of $\alpha$, $\delta$, and $h/J^{\rm NN}_\alpha$, the results are averaged over $1000$ independent disorder realizations. The gray vertical lines mark the points at $\alpha{=}0,1,2$. Results are shown for $N{=}14$ spins for the clean system and $N{=}10$ for the disordered system with (a) $h/J^{\rm NN}_\alpha{=}0.0$, (b) $h/J^{\rm NN}_\alpha{=}0.1$, (c) $h/J^{\rm NN}_\alpha{=}0.3$, (d) $h/J^{\rm NN}_\alpha{=}0.5$, (e) $h/J^{\rm NN}_\alpha{=}0.8$, (f) $h/J^{\rm NN}_\alpha{=}1.0$, (g) $h/J^{\rm NN}_\alpha{=}1.5$, (h) $h/J^{\rm NN}_\alpha{=}2.0$, and (i) $h/J^{\rm NN}_\alpha{=}4.0$.}
    \label{fig: Level stat spin 1}
\end{figure*}

However, this universality of the ground state across short-range interactions breaks down for higher spins. For spin-$1$, compared to the spin-$1/2$ case, the overlap of the ground states with that of the ground state at $\alpha{=}2$ deviates rapidly as $\alpha$ is tuned away from $2$ [see Fig.~\ref{fig: HS overlap spin 1}]. This indicates that the change in ground state physics as a function of $\alpha$ is more pronounced for higher spins. Analogous to the spin-$1/2$ case, the addition of a three-spin interaction term to the Hamiltonian results in the ground state at the $\alpha{=}2$ point assuming a Pfaffian form, similar to the bosonic Moore-Read fractional quantum Hall state~\cite{Moore1991} at filling fraction $\nu{=}1$~\cite{Greiter2011}. Note that, unlike the Laughlin state, the Moore-Read state is not the highest-density exact zero-energy state of a two-body interaction, which sheds light on the lack of universality in the ground-state physics of the two-spin interaction generalized Haldane-Shastry Hamiltonian for spin-$1$. Nevertheless, deep inside the bulk, the properties of the spin-$1$ chain remain similar to those of the spin-$1/2$ system, as is evident from the level statistics [see Fig.~\ref{fig: Level stat spin 1}]. The results for the spin-$1$ system were obtained using QuSpin~\cite{Quspin1, Quspin2}.

\section{Computation of the full spectrum of the Haldane-Shastry model}
\label{app: HS_spectrum}
In this appendix, we outline two methods to construct energies for all the eigenstates of the Haldane-Shastry model. In Appendix~\ref{app: HS_spectrum_YT}, we outline the method of Ref.~\cite{Greiter2007} to obtain the spectrum of the HS model using the standard Young tableaux representation of the $SU(2)$ group. In Appendix~\ref{app: HS_spectrum_spinon_states}, we enumerate the distinct spinon states of the HS model, which yields a very loose upper bound on the number of unique eigenstates in the model. 

\subsection{Using Young tableaux}
\label{app: HS_spectrum_YT}
\begin{figure*}[htbp!]
    \centering
    \includegraphics[width=1\linewidth]{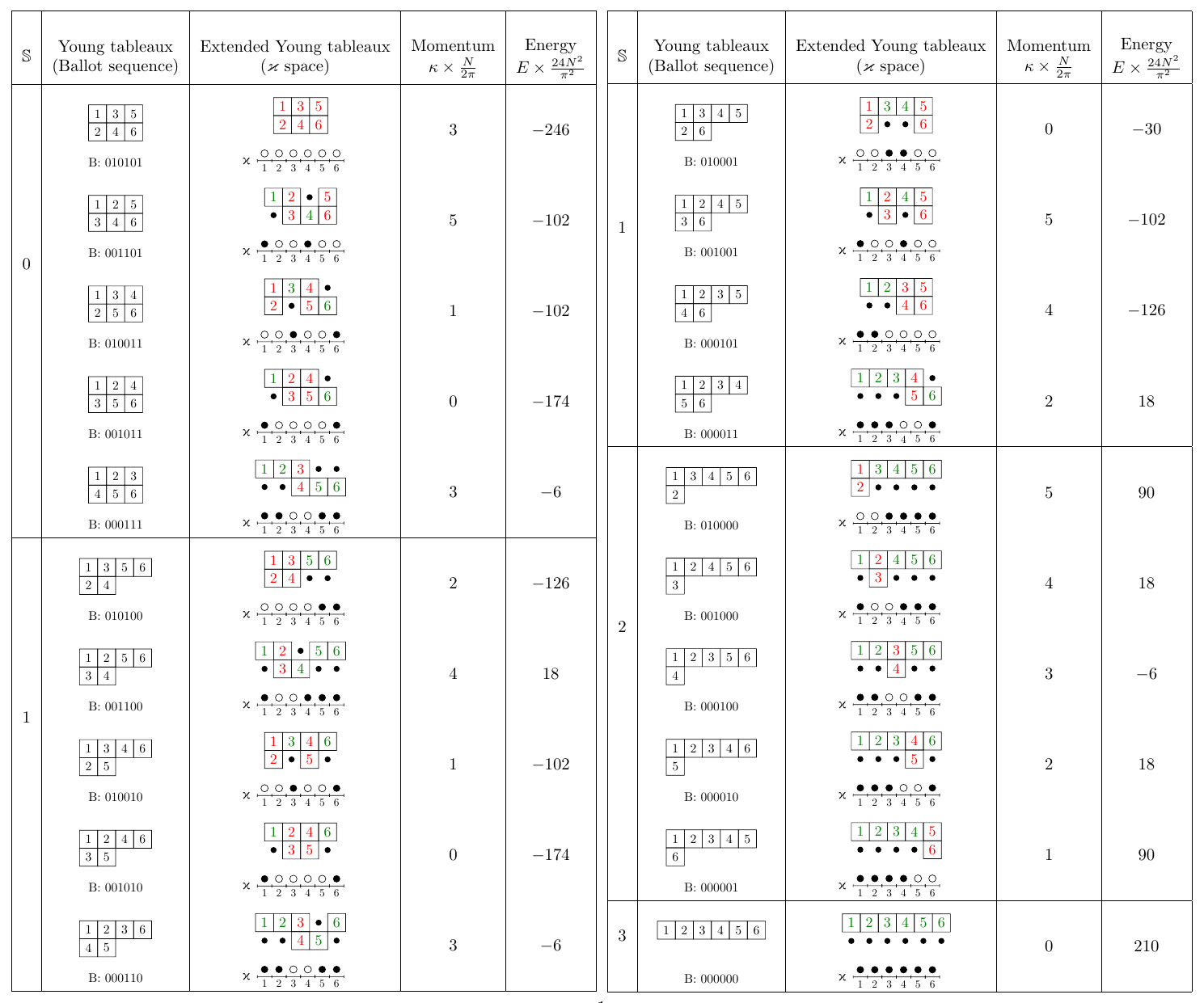}
    \caption{Pictorial illustration of all of the eigenstates of the Haldane-Shastry model for the zero magnetization sector of the $N{=}6$ spin-$1/2$ particles. The eigenstates are obtained using Young tableaux, and the spinons are shown by black dots, and their momentum quantum number $\varkappa$ is indicated. The first column shows the total spin $\vec{S}_{\rm tot}$ quantum number, $\mathbb{S}$. The second column shows the standard Young tableaux $SU(2)$ irreducible representations of $\mathbb{S}$, along with their bit-string representation. The third column gives the extended Young tableaux, which are obtained after the pairing of consecutive numbers in the top and bottom rows, and their corresponding $\varkappa$ space representation with spinons and their holes are shown as black and white dots. The second-last and last columns show the total momenta [see Eq.~\eqref{eq: many_spinon_p}] and energy [see Eq.~\eqref{eq: many_spinon_E}] of the eigenstates. }
    \label{fig: Young_tableaux}
\end{figure*}

A standard Young tableaux is an arrangement of $N$ numbered boxes with numbers ranging from $1,2,{\cdots}, N$ such that the numbers assigned to the boxes within each row increase from left to right and within each column increase from top to bottom. Each tableau should be symmetrized over all boxes in a given row and antisymmetrized over all boxes in a given column. Thus, for $SU(2)$ spins, it is not possible to have more than two boxes stacked vertically on top of each other since there are only two available spin states. Each tableau corresponds to a total spin quantum number $\vec{S}_{\rm tot}$, $\mathbb{S}$, and thus represents an eigenstate with that $\mathbb{S}$. To obtain the spectrum of the HS model, the standard tableaux are transformed into extended tableaux by rearranging the boxes in each row to make pairs such that the boxes with numbers $n$ and $n{+}1$ are in the same column, and the remaining unpaired boxes represent a spinon, with the number in the box being the momentum quantum number $\varkappa$ of that spinon. A pictorial illustration of this method to obtain $\varkappa$ is presented in Fig. \ref{fig: Young_tableaux}. The momentum $\kappa$ corresponding to a spinon with momentum quantum number $\varkappa$ is given by
\begin{equation}
    \kappa = \frac{\pi}{N} \left(\varkappa - \frac{1}{2} \right).
    \label{eq: spinon_momentum}
\end{equation}
A spinon with momentum $\kappa$ has an energy given by the following quadratic dispersion relation,
\begin{equation}
    \epsilon(\kappa) = \frac{1}{2} \kappa (\pi - \kappa) + \frac{\pi^2}{8N^2}.
    \label{eq: spion_dispersion}
\end{equation}
As the spinons in the HS model are free (implied by the quadratic dispersion), the total momentum and energy of a state with $l$ spinons is simply given by the sum of their individual values, i.e., 
\begin{equation}
    \kappa= \kappa_0 + \sum_{p=1}^{l} \kappa_p,
    \label{eq: many_spinon_p}
\end{equation}
and
\begin{equation}
    \quad E = E_0 + \sum_{p=1}^{l} \epsilon(\kappa_p).
    \label{eq: many_spinon_E}
\end{equation}
Here, $\kappa_0$ and $E_0$ are the ground state momentum and energy given by~\cite{Greiter2011}
\begin{equation}
    \kappa_0 = -\frac{\pi}{2} N, \quad E_0 = -\frac{\pi^2}{24}\left(N + \frac{5}{N}\right).
    \label{eq: P_E_Ground_state}
\end{equation}
Each Young tableau corresponds to an irreducible SU(2) representation,  i.e., to a multiplet with total spin $\mathbb{S}$. For even $N$, each multiplet contributes exactly one state with $m = 0$, so the total number of tableaux equals the dimension of the zero-magnetization sector, $\binom{N}{N/2}$. Thus, constructing the extended tableaux for all allowed shapes corresponding to $N$ spins yields the full spin spectrum.

The standard Young tableaux representation for $SU(n)$ spins with $N$ total spins can be represented as a Ballot sequence  \cite{AddarioBerry2008} of length $N$ with numbers $1,2,\cdots,n$, where each number represents the row at which the index number lies in the tableaux. There are $n^N$ possible sequences with the above setup. However, a valid Ballot sequence is obtained only when at any index in the sequence, the total number of occurrences of $i$ is greater than that of $j$ $\forall \,\, i{<}j$, i.e., the smaller number always dominates the sequence. This is the same condition for the standard Young tableaux, where the smaller number occupies the top rows. For $SU(2)$, as there are only two rows, the sequence can be represented as a bit-string with $0$ representing the first row and $1$ representing the second row (see Fig. \ref{fig: Young_tableaux}). As bits are native in the computing architecture, this representation allows for faster computation. Next, we show how the unique states in the HS model can be enumerated by counting its distinct spinon states.

\subsection{Using distinct spinon states} 
\label{app: HS_spectrum_spinon_states}
\begin{figure}[htbp!]
    \centering
    \includegraphics[width=1.0\linewidth]{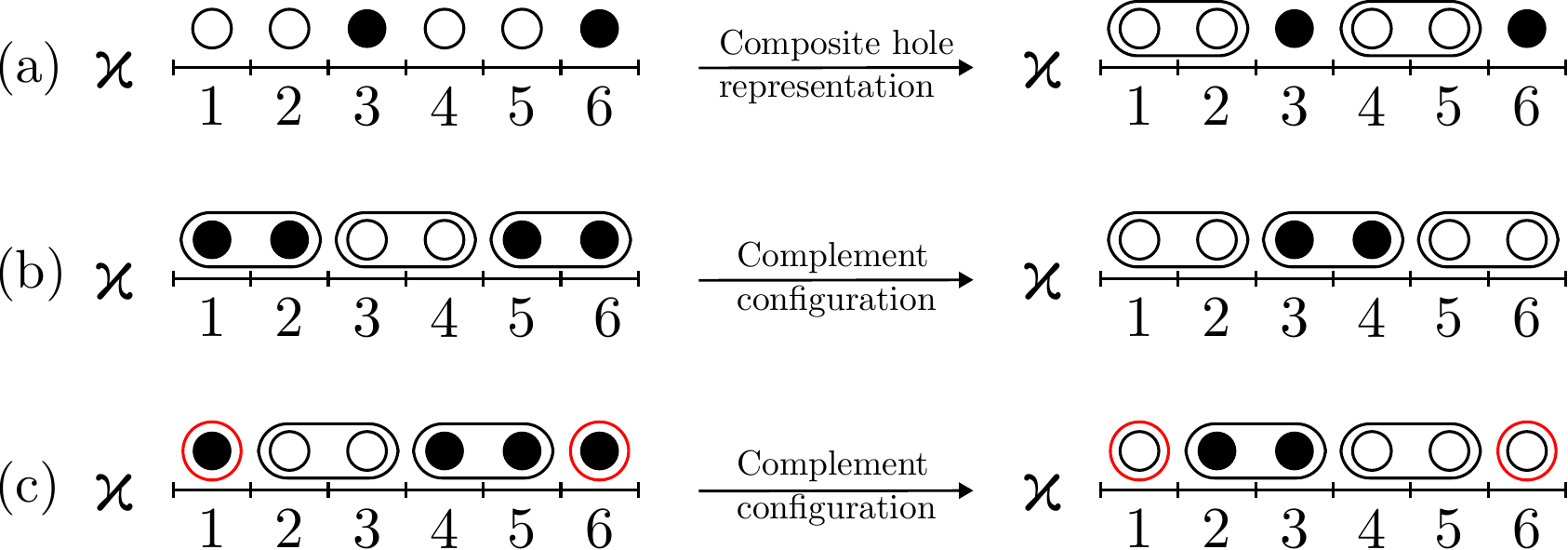}
    \caption{Pictorial illustration of spinons (black dots) and holes (white dots) in $\varkappa$ space in the Haldane-Shastry model for $N{=}6$ spin-$1/2$ particles. (a) The pairing of consecutive numbers in the extended Young tableaux forces the holes to appear in pairs occupying consecutive $\varkappa$, which behaves like a single entity, we refer to it as a composite hole and represent it by a black envelope surrounding the two holes; (b,c) Configurations in $\varkappa$ space along with its complement configuration obtained by interchanging spinons and holes, (b) A configuration whose complement exists and has particle hole like symmetry, (c) A configuration whose complement does not exist since in the complement the holes do not appear in pairs, i.e., there are single holes that are not part of composite holes (illustrated with a red envelope). Thus, the original configuration illustrated on the left-hand side of panel (c) does not have a hole partner.}
    \label{fig: k_space_illustrations}
\end{figure}
The extended Young tableaux, discussed in Appendix~\ref{app: HS_spectrum_YT}, admit a representation in the momentum index $\varkappa$ space, where the spinons occupy their corresponding $\varkappa$ and the remaining empty  $\varkappa$ slots are interpreted as occupied by spinon holes. In this picture, all the distinct configurations in the $\varkappa$ space are distinct spinon states of the HS model. Owing to the pairing of consecutive numbers in the construction of the extended Young tableaux, the holes always appear in pairs occupying consecutive $\varkappa$. Each such pair can be viewed as a single entity, which we refer to as a composite hole (see Fig. \ref{fig: k_space_illustrations}). The number of distinct configurations with $l$ composite holes is $\binom{N-l}{l}$, since it is sufficient to specify just the momentum index of the left hole, since its partner, right hole's momentum gets automatically specified due to them occupying consecutive $\varkappa$ indices (In other words, the momentum index of the right hole in the the composite hole is redundant.) Thus, since in all there can be atmost $N/2$ composite holes, the total number of distinct spinon states $\mathcal{N}_s$ is
\begin{equation}
\mathcal{N}_s = \sum_{0\le l \le N/2} \binom{N-l}{l} = F_{N+1},
\label{eq: num_spinon_states}
\end{equation}
where $F_{N{+}1}$ is the $(N{+}1)^{\text{th}}$ Fibbonacci number. It turns out that the result stated in Eq.~\eqref{eq: num_spinon_states} broadly applies to any Yangian invariant $SU(2)$ spin model~\cite{Finkel_Lopez_2015_Yangian_Fibonacci}. Since $F_{N{+}1}$ grows slower than $\binom{N}{N/2}$ and the spinons and composite holes can also be represented by bits, this representation allows us to construct the unique spinon eigenstates of the HS model for systems larger than the ones accessible to the Young tableaux. It is important to note that the total number of distinct spinon states given in Eq.~\eqref{eq: num_spinon_states}, which grows exponentially in $N$, is way more than the total number of unique states given in Eq.~\eqref{eq: upper_bound_N_E_HS}, which only grows polynomially in $N$. This is because distinct spinon states can map to the same energy owing to the dispersion relation given in Eq.~\eqref{eq: spion_dispersion}.
\begin{figure*}[tb]
    \centering
    \includegraphics[width=1\linewidth]{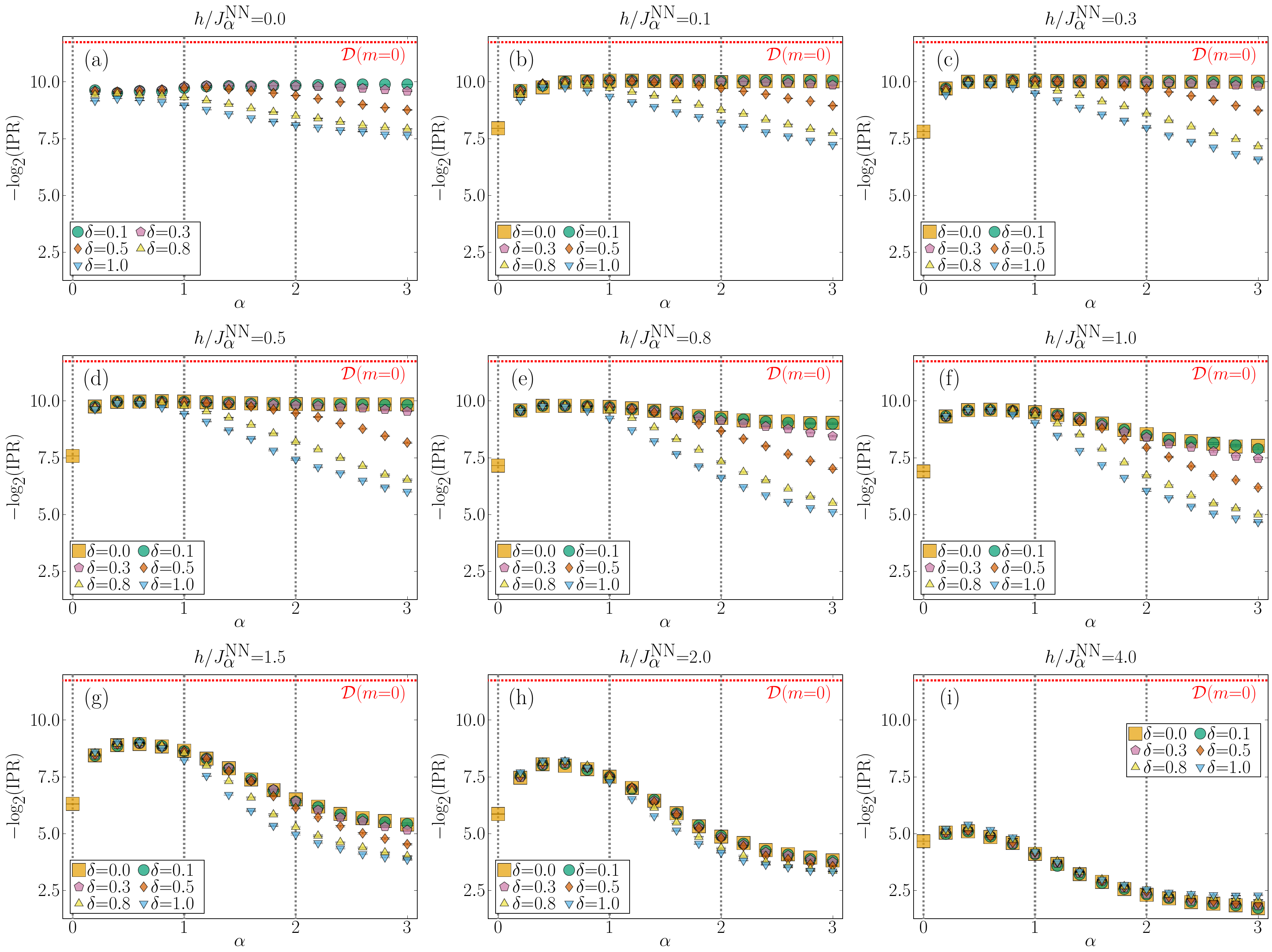}
    \caption{Disorder averaged $\mathrm{IPR}$ for the middle $10\%$ of eigenstates in the spectrum of the Hamiltonian [see Eq.~\eqref{eq: z_h_dis_HS_gen_Hamiltonian}] in $-\mathrm{log_2}$ scale as a function of the interaction parameter $\alpha$ for different values of position disorder strengths $\delta$ and magnetic field disorder strengths $h/J^{\rm NN}_\alpha$ in the zero magnetization sector. For each value of $\alpha$, $\delta$, and $h/J^{\rm NN}_\alpha$, the results are averaged over $1000$ independent disorder realizations. The gray vertical lines mark the points at $\alpha{=}0,1,2$. The red horizontal line marks the dimension of the zero magnetization sector. Results are shown for $N{=}14$ spins with (a) $h/J^{\rm NN}_\alpha{=}0.0$; (b) $h/J^{\rm NN}_\alpha{=}0.1$; (c) $h/J^{\rm NN}_\alpha{=}0.3$; (d) $h/J^{\rm NN}_\alpha{=}0.5$; (e) $h/J^{\rm NN}_\alpha{=}0.8$; (f) $h/J^{\rm NN}_\alpha{=}1.0$; (g) $h/J^{\rm NN}_\alpha{=}1.5$; (h) $h/J^{\rm NN}_\alpha{=}2.0$; (i) $h/J^{\rm NN}_\alpha{=}4.0$. The $\mathrm{IPR}$ demonstrates behavior consistent with the level statistics for all $h/J^{\rm NN}_\alpha{\neq}0.0$.}
    \label{fig: ipr_plots}
\end{figure*}

\section{Relating certain configurations in the Haldane-Shastry model via an emergent particle-hole symmetry in the thermodynamic limit}
\label{app: emergent_PHS_HS}

In this appendix, we demonstrate an emergent particle-hole-like symmetry that relates certain configurations in the HS model in the thermodynamic limit using the formalism developed in Appendix~\ref{app: HS_spectrum}. The number of such configurations is exponentially smaller than the total number of configurations; thus, this result does not imply that the HS model in its entirety has a particle-hole-like symmetry. In the thermodynamic limit, Eqs.~\eqref{eq: spinon_momentum},~\eqref{eq: spion_dispersion} and \eqref{eq: P_E_Ground_state} can be approximated as
\begin{equation}
    \kappa_{N\to \infty} =\lim_{N\to \infty} \frac{\pi}{N} \left( \varkappa - \frac{1}{2} \right) \approx  \frac{\pi}{N} \varkappa ,
    \label{eq: spinon_momentum_thermo}
\end{equation}
\begin{eqnarray}  
    \epsilon(\kappa)_{N\to \infty} &=& \lim_{N\to \infty} \frac{1}{2} \kappa_{N\to \infty} (\pi - \kappa_{N\to \infty}) + \frac{\pi^2}{8N^2} \nonumber\\ 
    &\approx& \frac{1}{2} \kappa_{N\to \infty} (\pi - \kappa_{N\to \infty}), \\
   \epsilon(\varkappa)_{N\to \infty} &=& \frac{\pi^2}{2N} \varkappa - \frac{\pi^2}{2N^2} \varkappa^2,
    \label{eq: spion_dispersion_thermo}
\end{eqnarray}
\begin{equation}
     (E_{0})_{N\to \infty} = \lim_{N\to \infty} -\frac{\pi^2}{24}\left(N + \frac{5}{N}\right) \approx -\frac{\pi^2}{24}N.
    \label{eq: E_Ground_state_thermo}
\end{equation}
In the $\varkappa$ space picture, the ground state is the state with $N/2$ composite holes. The maximum energy state is the state without holes, and the energy of that state is
\begin{eqnarray}
\label{eq: E_max_state_thermo}
     (E_{\rm max})_{N\to \infty} &\approx& -\frac{\pi^2}{24}N + \sum_{1 \le \varkappa \le N}\epsilon(\varkappa)_{N\to \infty} \nonumber \\   
     &=& -\frac{\pi^2}{24}N + \sum_{1 \le \varkappa \le N}\frac{\pi^2}{2N} \varkappa - \frac{\pi^2}{2N^2} \varkappa^2 \nonumber \\ 
     &=& \frac{\pi^2}{24}N-\frac{\pi^2}{12N} \approx \frac{\pi^2}{24}N,
\end{eqnarray}
which is $-(E_{0})_{N\to \infty}$. This correspondence suggests there could be a spectral reflection or particle-hole-like symmetry between spinons and their holes in the HS model in the thermodynamic limit. The energy of the state with $2l$ spinons with configuration $\{\varkappa\}$ in the $\varkappa$ space is
\begin{equation}
    E(\{\varkappa\})_{N\to \infty} \approx -\frac{\pi^2}{24}N + \sum_{\{\varkappa\}} \epsilon(\varkappa)_{N\to \infty}.
    \label{eq: E_2l_spinon_thermo}
\end{equation}
Similarly, the energy of the state with $2l$ spinon holes with configuration $\{\bar{\varkappa}\}$, where $\{\bar{\varkappa}\}$ is obtained from $\{\varkappa\}$ by interchanging spinons and holes, is given by 
\begin{eqnarray}
    E(\{\bar{\varkappa}\})_{N\to \infty} &\approx& -\frac{\pi^2}{24}N + \sum_{\{\bar{\varkappa}\}} \epsilon(\varkappa)_{N\to \infty} \\ \nonumber
    &=& -\frac{\pi^2}{24}N + \sum_{1 \le \varkappa \le N}\epsilon(\varkappa)_{N\to \infty} -\sum_{\{\varkappa\}} \epsilon(\varkappa)_{N\to \infty}\\ \nonumber
    &=&  \frac{\pi^2}{24}N - \sum_{\{\varkappa\}} \epsilon(\varkappa)_{N\to \infty} \\ \nonumber
    &=& -E(\{\varkappa\})_{N\to \infty}~[{\rm see~Eq.~\eqref{eq: E_2l_spinon_thermo}}],
    \label{eq: E_2l_hole_thermo}
\end{eqnarray}
which implies that in the HS model in the thermodynamic limit, certain complementary configurations can be related by a particle-hole-like symmetry. Since the holes always appear in pairs, such complement states exist only when the spinons also appear in pairs (see Fig. \ref{fig: k_space_illustrations}). This pairing results in effectively halving the degrees of freedom in the $\varkappa$ space, so the total number of states that are made from such pairs is $2^{N/2}$. In this subspace, there is an emergent particle-hole symmetry in the thermodynamic limit in the HS model due to vanishing $\mathcal{O}(1/N)$ corrections, which allows relating a configuration with its spectrum-reflected partner. Nevertheless, the number of such states is exponentially smaller than the total number of distinct spinon states, which is $F_{N{+}1}$ [see Eq.~\eqref{eq: num_spinon_states}], and thereby does not imply the existence of a particle-hole-like symmetry in the entire HS model. 

\section{Inverse Participation Ratio}
\label{app: ipr}

The inverse participation ratio (IPR) serves as a diagnostic to measure the degree of localization of a state within a specific basis. In the $S^z$ basis, it is defined as
\begin{equation}
    \mathrm{IPR} = \sum_{i=1}^{\mathcal{D}} |\langle i | \psi \rangle|^4,
    \label{eq: IPR_def}
\end{equation}
where $|\psi\rangle$ is a normalized many-body eigenstate, $|i\rangle$ is the $i^{\mathrm{th}}$ basis state in the $S^z$ basis, and $\mathcal{D}$ is the dimension of the Hilbert space. For a completely localized state, the $\mathrm{IPR}$ is $1$, as the wavefunction has non-zero amplitude in only a single basis state. Conversely, for a fully delocalized state that is spread uniformly across all the basis states, the $\mathrm{IPR}$ scales as $1/\mathcal{D}$, and goes to zero in the thermodynamic limit.

In the position disordered model of Eq.~\eqref{eq: z_dis_HS_gen_Hamiltonian}, the states are not localized, as expected due to the presence of $SU(2)$ symmetry; however, the $\mathrm{IPR}$ exhibits a dependence on $\delta$ [see Fig.~\ref{fig: ipr_plots}(a)], which is in contrast with the behavior of the level statistics. For the model with both position and magnetic field disorders [see Eq.~\eqref{eq: z_h_dis_HS_gen_Hamiltonian}], the $\mathrm{IPR}$ demonstrates behavior consistent with the level statistics, further corroborating our results [see Fig.~\ref{fig: ipr_plots}(b)-\ref{fig: ipr_plots}(i)] with $\alpha{=}0$ being an exception. At $\alpha{=}0$, the states remain delocalized even though the level statistics exhibit a Poisson distribution.

\bibliography{references}

@misc{bai2026bootstrappingsymmetriesquantummanybody,
      title={Bootstrapping Symmetries in Quantum Many-Body Systems from the Cross Spectral Form Factor}, 
      author={Chen Bai and Zihan Zhou and Bastien Lapierre and Shinsei Ryu},
      year={2026},
      eprint={2604.01296},
      archivePrefix={arXiv},
      primaryClass={quant-ph},
      url={https://arxiv.org/abs/2604.01296}, 
}

@misc{odea2024levelstatisticsdetectgeneralized,
      title={Level statistics detect generalized symmetries}, 
      author={Nicholas O'Dea},
      year={2024},
      eprint={2406.03983},
      archivePrefix={arXiv},
      primaryClass={cond-mat.stat-mech},
      url={https://arxiv.org/abs/2406.03983}, 
}

@article{Haldane88,
  title = {Exact {Jastrow}-{Gutzwiller} resonating-valence-bond ground state of the spin-$\frac{1}{2}$ antiferromagnetic {Heisenberg} chain with 1/${\mathrm{r}}^{2}$ exchange},
  author = {Haldane, F. D. M.},
  journal = {Phys. Rev. Lett.},
  volume = {60},
  issue = {7},
  pages = {635--638},
  numpages = {0},
  year = {1988},
  month = {Feb},
  publisher = {American Physical Society},
  doi = {10.1103/PhysRevLett.60.635},
  url = {https://link.aps.org/doi/10.1103/PhysRevLett.60.635}
}

@article{he2026spectral,
  title = {Spectral decimation of quantum many-body {Hamiltonians}},
  author = {He, Feng and Hutsalyuk, Arthur and Mussardo, Giuseppe and Stampiggi, Andrea},
  journal = {Phys. Rev. B},
  volume = {113},
  issue = {24},
  pages = {245121},
  numpages = {19},
  year = {2026},
  month = {Jun},
  publisher = {American Physical Society},
  doi = {10.1103/4dct-rs4q},
  url = {https://link.aps.org/doi/10.1103/4dct-rs4q}
}

@article{Shastry88,
  title = {Exact solution of an {S=1/2} {Heisenberg} antiferromagnetic chain with long-ranged interactions},
  author = {Shastry, B. Sriram},
  journal = {Phys. Rev. Lett.},
  volume = {60},
  issue = {7},
  pages = {639--642},
  numpages = {0},
  year = {1988},
  month = {Feb},
  publisher = {American Physical Society},
  doi = {10.1103/PhysRevLett.60.639},
  url = {https://link.aps.org/doi/10.1103/PhysRevLett.60.639}
}

@book{Greiter2011,
  author = {Greiter, Martin},
  title = {Mapping of Parent Hamiltonians},
  subtitle = {From Abelian and non-Abelian Quantum Hall States to Exact Models of Critical Spin Chains},
  series = {Springer Tracts in Modern Physics},
  volume = {248}, 
  publisher = {Springer Berlin, Heidelberg},
  year = {2011},
  doi = {10.1007/978-3-642-24384-4},
  isbn = {978-3-642-24383-7},
  pages = {XIV, 194},
  url = {https://doi.org/10.1007/978-3-642-24384-4}
}

@article{Laughlin83,
  title = {Anomalous {Quantum} {Hall} Effect: An Incompressible Quantum Fluid with Fractionally Charged Excitations},
  author = {Laughlin, R. B.},
  journal = {Phys. Rev. Lett.},
  volume = {50},
  issue = {18},
  pages = {1395--1398},
  numpages = {0},
  year = {1983},
  month = {May},
  publisher = {American Physical Society},
  doi = {10.1103/PhysRevLett.50.1395},
  url = {https://link.aps.org/doi/10.1103/PhysRevLett.50.1395}
}

@article{Yangian_Haldane_et_al_92,
  title = {Yangian symmetry of integrable quantum chains with long-range interactions and a new description of states in conformal field theory},
  author = {Haldane, F. D. M. and Ha, Z. N. C. and Talstra, J. C. and Bernard, D. and Pasquier, V.},
  journal = {Phys. Rev. Lett.},
  volume = {69},
  issue = {14},
  pages = {2021--2025},
  numpages = {0},
  year = {1992},
  month = {Oct},
  publisher = {American Physical Society},
  doi = {10.1103/PhysRevLett.69.2021},
  url = {https://link.aps.org/doi/10.1103/PhysRevLett.69.2021}

}

@misc{talstra1995integrability_HS,
      title={Integrability and Applications of the Exactly-Solvable {Haldane-Shastry} One-Dimensional Quantum Spin Chain}, 
      author={Johan Cornelis Talstra},
      year={1995},
      archivePrefix={arXiv},
      primaryClass={cond-mat},
      url={https://arxiv.org/abs/cond-mat/9509178}, 
}

@article{Squeezed_strings_and_Yangian_Haldane_et_al93,
  title = {Squeezed strings and {Yangian} symmetry of the {Heisenberg} chain with long-range interaction},
  author = {Ha, Z. N. C. and Haldane, F. D. M.},
  journal = {Phys. Rev. B},
  volume = {47},
  issue = {19},
  pages = {12459--12469},
  numpages = {0},
  year = {1993},
  month = {May},
  publisher = {American Physical Society},
  doi = {10.1103/PhysRevB.47.12459},
  url = {https://link.aps.org/doi/10.1103/PhysRevB.47.12459}
}

@inbook{Hoft_algebra_DRINFEL,
    author = {V. G. Drinfel'D},
    title = {HOPF ALGEBRAS AND THE QUANTUM YANG-BAXTER EQUATION},
    booktitle = {Yang-Baxter Equation in Integrable Systems},
    publisher = {World Scientific},
    pages = {264-268},
    year={1990},
    doi = {10.1142/9789812798336_0013},
    URL = {https://worldscientific.com/doi/abs/10.1142/9789812798336\textunderscore0013},
}

@book{chari1995guide_QG,
  title={A Guide to Quantum Groups},
  author={Chari, V. and Pressley, A.N.},
  isbn={9780521558846},
  lccn={94241658},
  url={https://www.cambridge.org/us/universitypress/subjects/mathematics/algebra/guide-quantum-groups},
  year={1995},
  publisher ={Cambridge University Press}
}

@article{Talstra_1995,
doi = {10.1088/0305-4470/28/8/027},
url = {https://dx.doi.org/10.1088/0305-4470/28/8/027},
year = {1995},
month = {apr},
publisher = {},
volume = {28},
number = {8},
pages = {2369},
author = {J C Talstra and F D M Haldane},
title = {Integrals of motion of the {Haldane-Shastry} model},
journal = {Journal of Physics A: Mathematical and General},
}

@article{Greiter2007,
  author = {Greiter, M. and Schuricht, D.},
  title = {Many-spinon states and the secret significance of {Young} tableaux},
  journal = {Physical Review Letters},
  volume = {98},
  number = {23},
  pages = {237202},
  year = {2007},
  doi = {10.1103/PhysRevLett.98.237202},
  url = {https://doi.org/10.1103/PhysRevLett.98.237202}
}

@article{Sandvik_2010,
    author = {Sandvik, Anders W.},
    title = {Computational Studies of Quantum Spin Systems},
    journal = {AIP Conference Proceedings},
    volume = {1297},
    number = {1},
    pages = {135-338},
    year = {2010},
    month = {11},
    issn = {0094-243X},
    doi = {10.1063/1.3518900},
    url = {https://doi.org/10.1063/1.3518900},
}

@article{Atas_Bogomolny_Giraud_Roux_2013, 
title={Distribution of the Ratio of Consecutive Level Spacings in Random Matrix Ensembles}, 
volume={110},
url={https://link.aps.org/doi/10.1103/PhysRevLett.110.084101}, 
DOI={10.1103/PhysRevLett.110.084101}, 
number={8}, 
journal={Physical Review Letters}, 
publisher={American Physical Society}, 
author={Atas, Y. Y. and Bogomolny, E. and Giraud, O. and Roux, G.},
year={2013},
month=feb, 
pages={084101} }

@article{Kudo_Deguchi_2003,
  title = {Unexpected non-{Wigner} behavior in level-spacing distributions of next-nearest-neighbor coupled $\mathrm{{X}{X}{Z}}$ spin chains},
  author = {Kudo, Kazue and Deguchi, Tetsuo},
  journal = {Phys. Rev. B},
  volume = {68},
  issue = {5},
  pages = {052510},
  numpages = {4},
  year = {2003},
  month = {Aug},
  publisher = {American Physical Society},
  doi = {10.1103/PhysRevB.68.052510},
  url = {https://link.aps.org/doi/10.1103/PhysRevB.68.052510}
}

@misc{Hallam25,
      title={Tunable quantum {Mpemba} effect in long-range interacting systems}, 
      author={Andrew Hallam and Matthew Yusuf and Aashish A. Clerk and Ivar Martin and Zlatko Papić},
      year={2025},
      eprint={2510.12875},
      archivePrefix={arXiv},
      primaryClass={quant-ph},
      url={https://arxiv.org/abs/2510.12875}, 
}

@article{Oganesyan_et.al_2007,
  title = {Localization of interacting fermions at high temperature},
  author = {Oganesyan, Vadim and Huse, David A.},
  journal = {Phys. Rev. B},
  volume = {75},
  issue = {15},
  pages = {155111},
  numpages = {5},
  year = {2007},
  month = {Apr},
  publisher = {American Physical Society},
  doi = {10.1103/PhysRevB.75.155111},
  url = {https://link.aps.org/doi/10.1103/PhysRevB.75.155111}
}

@article{Shriya_Pai_2017,
  title = {Disordered {Haldane-Shastry} model},
  author = {Pai, Shriya and Srivatsa, N. S. and Nielsen, Anne E. B.},
  journal = {Phys. Rev. B},
  volume = {102},
  issue = {3},
  pages = {035117},
  numpages = {6},
  year = {2020},
  month = {Jul},
  publisher = {American Physical Society},
  doi = {10.1103/PhysRevB.102.035117},
  url = {https://link.aps.org/doi/10.1103/PhysRevB.102.035117}
}

@Article{Trugman85,
  Title                    = {Exact results for the fractional quantum {Hall} effect with general interactions},
  Author                   = {Trugman, S. A. and Kivelson, S.},
  Journal                  = {Phys. Rev. B},
  Year                     = {1985},

  Month                    = {Apr},
  Pages                    = {5280--5284},
  Volume                   = {31},
  Doi                      = {10.1103/PhysRevB.31.5280},
  Issue                    = {8},
  Numpages                 = {5},
  Publisher                = {American Physical Society},
  Url                      = {http://link.aps.org/doi/10.1103/PhysRevB.31.5280}
}

@Article{Haldane83,
  Title                    = {Fractional Quantization of the {Hall} Effect: A Hierarchy of Incompressible Quantum Fluid States},
  Author                   = {Haldane, F. D. M.},
  Journal                  = {Phys. Rev. Lett.},
  Year                     = {1983},

  Month                    = {Aug},
  Pages                    = {605--608},
  Volume                   = {51},
  Doi                      = {10.1103/PhysRevLett.51.605},
  Issue                    = {7},
  Numpages                 = {0},
  Publisher                = {American Physical Society},
  Url                      = {http://link.aps.org/doi/10.1103/PhysRevLett.51.605}
}

@article{Nielsen_2011,
doi = {10.1088/1742-5468/2011/11/P11014},
url = {https://dx.doi.org/10.1088/1742-5468/2011/11/P11014},
year = {2011},
month = {nov},
publisher = {},
volume = {2011},
number = {11},
pages = {P11014},
author = {Nielsen, Anne E B and Ignacio Cirac, J and Sierra, Germán},
title = {Quantum spin {Hamiltonians} for the {$SU(2)_k$} {WZW} model},
journal = {Journal of Statistical Mechanics: Theory and Experiment},
}

@article{Cirac_Sierra_2010,
  title = {Infinite matrix product states, conformal field theory, and the {Haldane-Shastry} model},
  author = {Cirac, J. Ignacio and Sierra, Germ\'an},
  journal = {Phys. Rev. B},
  volume = {81},
  issue = {10},
  pages = {104431},
  numpages = {4},
  year = {2010},
  month = {Mar},
  publisher = {American Physical Society},
  doi = {10.1103/PhysRevB.81.104431},
  url = {https://link.aps.org/doi/10.1103/PhysRevB.81.104431}
}

@book{Haake2018Quantum,
  title        = {Quantum Signatures of Chaos},
  author       = {Haake, Fritz and Gnutzmann, Sven and Ku{\'s}, Marek},
  year         = {2018},
  publisher    = {Springer International Publishing},
  address      = {Cham},
  doi          = {10.1007/978-3-319-97580-1},
  isbn         = {978-3-319-97579-5, 978-3-319-97580-1},
}

@article{Avishai2002,
  title = {Level statistics in a {Heisenberg} chain with random magnetic field},
  author = {Avishai, Y. and Richert, J. and Berkovits, R.},
  journal = {Phys. Rev. B},
  volume = {66},
  issue = {5},
  pages = {052416},
  numpages = {4},
  year = {2002},
  month = {Aug},
  publisher = {American Physical Society},
  doi = {10.1103/PhysRevB.66.052416},
  url = {https://link.aps.org/doi/10.1103/PhysRevB.66.052416}
}

@book{Kashiwara2002,
  title        = {MathPhys {Odyssey} 2001: {Integrable} Models and Beyond --- {In} Honor of {Barry M. McCoy}},
  editor       = {Masaki Kashiwara and Tetsuji Miwa},
  series       = {Progress in Mathematical Physics},
  volume       = {23},
  publisher    = {Birkhäuser Boston},
  address      = {Boston, MA},
  year         = {2002},
  doi          = {10.1007/978-1-4612-0087-1},
  isbn         = {978-0-8176-4260-0},
  eisbn        = {978-1-4612-0087-1},
  series_issn  = {1544-9998},
  series_eissn = {2197-1846},
  note         = {Springer Book Archive},
  pages        = {XI + 476},
  url          = {https://doi.org/10.1007/978-1-4612-0087-1}
}

@article{Deguchi2003,
author = {Kudo ,Kazue and Deguchi ,Tetsuo},
title = {Branches in the Spectral Flow of the Inhomogeneous Transfer Matrix for the {XXZ} Spin Chain},
journal = {Journal of the Physical Society of Japan},
volume = {72},
number = {7},
pages = {1599-1602},
year = {2003},
doi = {10.1143/JPSJ.72.1599},
URL = {https://doi.org/10.1143/JPSJ.72.1599},
eprint = { https://doi.org/10.1143/JPSJ.72.1599}
}

@article{Deutsch91,
  title = {Quantum statistical mechanics in a closed system},
  author = {Deutsch, J. M.},
  journal = {Phys. Rev. A},
  volume = {43},
  issue = {4},
  pages = {2046--2049},
  numpages = {0},
  year = {1991},
  month = {Feb},
  publisher = {American Physical Society},
  doi = {10.1103/PhysRevA.43.2046},
  url = {https://link.aps.org/doi/10.1103/PhysRevA.43.2046}
}

@article{Srednicki94,
  title = {Chaos and quantum thermalization},
  author = {Srednicki, Mark},
  journal = {Phys. Rev. E},
  volume = {50},
  issue = {2},
  pages = {888--901},
  numpages = {0},
  year = {1994},
  month = {Aug},
  publisher = {American Physical Society},
  doi = {10.1103/PhysRevE.50.888},
  url = {https://link.aps.org/doi/10.1103/PhysRevE.50.888}
}

@article{Rigol2008,
  author = {Rigol, Marcos and Dunjko, Vanja and Olshanii, Maxim},
  title = {Thermalization and its mechanism for generic isolated quantum systems},
  journal = {Nature},
  volume = {452},
  issue = {7189},
  pages = {854--858},
  year = {2008},
  month = {April},
  doi = {10.1038/nature06838},
  url = {https://doi.org/10.1038/nature06838},
  issn = {1476-4687},
}

@article{Polkovnikov_2011,
  title = {Colloquium: {Nonequilibrium} dynamics of closed interacting quantum systems},
  author = {Polkovnikov, Anatoli and Sengupta, Krishnendu and Silva, Alessandro and Vengalattore, Mukund},
  journal = {Rev. Mod. Phys.},
  volume = {83},
  issue = {3},
  pages = {863--883},
  numpages = {0},
  year = {2011},
  month = {Aug},
  publisher = {American Physical Society},
  doi = {10.1103/RevModPhys.83.863},
  url = {https://link.aps.org/doi/10.1103/RevModPhys.83.863}
}

@article{Rigol_2012,
  title = {Alternatives to Eigenstate Thermalization},
  author = {Rigol, Marcos and Srednicki, Mark},
  journal = {Phys. Rev. Lett.},
  volume = {108},
  issue = {11},
  pages = {110601},
  numpages = {5},
  year = {2012},
  month = {Mar},
  publisher = {American Physical Society},
  doi = {10.1103/PhysRevLett.108.110601},
  url = {https://link.aps.org/doi/10.1103/PhysRevLett.108.110601}
}

@article{Kim_2014,
  title = {Testing whether all eigenstates obey the eigenstate thermalization hypothesis},
  author = {Kim, Hyungwon and Ikeda, Tatsuhiko N. and Huse, David A.},
  journal = {Phys. Rev. E},
  volume = {90},
  issue = {5},
  pages = {052105},
  numpages = {8},
  year = {2014},
  month = {Nov},
  publisher = {American Physical Society},
  doi = {10.1103/PhysRevE.90.052105},
  url = {https://link.aps.org/doi/10.1103/PhysRevE.90.052105}
}

@article{Deutsch_2018,
doi = {10.1088/1361-6633/aac9f1},
url = {https://doi.org/10.1088/1361-6633/aac9f1},
year = {2018},
month = {jul},
publisher = {IOP Publishing},
volume = {81},
number = {8},
pages = {082001},
author = {Deutsch, Joshua M},
title = {Eigenstate thermalization hypothesis},
journal = {Reports on Progress in Physics},
}

@article{Ueda2020,
  author = {Ueda, Masahito},
  title = {Quantum equilibration, thermalization and prethermalization in ultracold atoms},
  journal = {Nature Reviews Physics},
  volume = {2},
  pages = {669--681},
  year = {2020},
  doi = {10.1038/s42254-020-0237-x},
  url = {https://doi.org/10.1038/s42254-020-0237-x}
}

@article{Anderson58,
  title = {Absence of Diffusion in Certain Random Lattices},
  author = {Anderson, P. W.},
  journal = {Phys. Rev.},
  volume = {109},
  issue = {5},
  pages = {1492--1505},
  numpages = {0},
  year = {1958},
  month = {Mar},
  publisher = {American Physical Society},
  doi = {10.1103/PhysRev.109.1492},
  url = {https://link.aps.org/doi/10.1103/PhysRev.109.1492}
}

@article{Lee85,
  title = {Disordered electronic systems},
  author = {Lee, Patrick A. and Ramakrishnan, T. V.},
  journal = {Rev. Mod. Phys.},
  volume = {57},
  issue = {2},
  pages = {287--337},
  numpages = {0},
  year = {1985},
  month = {Apr},
  publisher = {American Physical Society},
  doi = {10.1103/RevModPhys.57.287},
  url = {https://link.aps.org/doi/10.1103/RevModPhys.57.287}
}

@article{Gornyi_2005,
  title = {Interacting Electrons in Disordered Wires: {Anderson} Localization and Low-$T$ Transport},
  author = {Gornyi, I. V. and Mirlin, A. D. and Polyakov, D. G.},
  journal = {Phys. Rev. Lett.},
  volume = {95},
  issue = {20},
  pages = {206603},
  numpages = {4},
  year = {2005},
  month = {Nov},
  publisher = {American Physical Society},
  doi = {10.1103/PhysRevLett.95.206603},
  url = {https://link.aps.org/doi/10.1103/PhysRevLett.95.206603}
}

@article{BASKO_2006,
title = {Metal–insulator transition in a weakly interacting many-electron system with localized single-particle states},
journal = {Annals of Physics},
volume = {321},
number = {5},
pages = {1126-1205},
year = {2006},
issn = {0003-4916},
doi = {https://doi.org/10.1016/j.aop.2005.11.014},
url = {https://www.sciencedirect.com/science/article/pii/S0003491605002630},
author = {D.M. Basko and I.L. Aleiner and B.L. Altshuler},
}

@article{Huse_2013,
  title = {Localization-protected quantum order},
  author = {Huse, David A. and Nandkishore, Rahul and Oganesyan, Vadim and Pal, Arijeet and Sondhi, S. L.},
  journal = {Phys. Rev. B},
  volume = {88},
  issue = {1},
  pages = {014206},
  numpages = {8},
  year = {2013},
  month = {Jul},
  publisher = {American Physical Society},
  doi = {10.1103/PhysRevB.88.014206},
  url = {https://link.aps.org/doi/10.1103/PhysRevB.88.014206}
}

@article{Nandkishore2015,
  author = {Nandkishore, Rahul and Huse, David A.},
  title = {Many-Body Localization and Thermalization in Quantum Statistical Mechanics},
  journal = {Annual Review of Condensed Matter Physics},
  volume = {6},
  pages = {15--38},
  year = {2015},
  doi = {10.1146/annurev-conmatphys-031214-014726},
  url = {https://doi.org/10.1146/annurev-conmatphys-031214-014726},
}

@article{Abanin_2019,
  title = {Colloquium: {Many-body} localization, thermalization, and entanglement},
  author = {Abanin, Dmitry A. and Altman, Ehud and Bloch, Immanuel and Serbyn, Maksym},
  journal = {Rev. Mod. Phys.},
  volume = {91},
  issue = {2},
  pages = {021001},
  numpages = {26},
  year = {2019},
  month = {May},
  publisher = {American Physical Society},
  doi = {10.1103/RevModPhys.91.021001},
  url = {https://link.aps.org/doi/10.1103/RevModPhys.91.021001}
}

@article{Kac1963,
  author = {Kac, Mark and Uhlenbeck, George E. and Hemmer, P. C.},
  title = {On the {van der Waals} Theory of the Vapor-Liquid Equilibrium. I. Discussion of a One-Dimensional Model},
  journal = {Journal of Mathematical Physics},
  volume = {4},
  issue = {2},
  pages = {216--228},
  year = {1963},
  month = {February},
  doi = {10.1063/1.1703946},
  url = {https://doi.org/10.1063/1.1703946}
}

@article{Wigner_1951,
title={On the statistical distribution of the widths and spacings of nuclear resonance levels},
volume={47},
DOI={10.1017/S0305004100027237},
number={4}, journal={Mathematical Proceedings of the Cambridge Philosophical Society},
author={Wigner, Eugene P.},
year={1951},
pages={790–798}
}

@article{Dyson1962,
  author = {Dyson, Freeman J.},
  title = {Statistical Theory of the Energy Levels of Complex Systems. I},
  journal = {J. Math. Phys.},
  volume = {3},
  issue = {1},
  pages = {140--156},
  year = {1962},
  doi = {10.1063/1.1703773},
  url = {https://doi.org/10.1063/1.1703773},
  publisher = {AIP}
}

@article{Pal_2010,
  title = {Many-body localization phase transition},
  author = {Pal, Arijeet and Huse, David A.},
  journal = {Phys. Rev. B},
  volume = {82},
  issue = {17},
  pages = {174411},
  numpages = {7},
  year = {2010},
  month = {Nov},
  publisher = {American Physical Society},
  doi = {10.1103/PhysRevB.82.174411},
  url = {https://link.aps.org/doi/10.1103/PhysRevB.82.174411}
}

@article{Berry77,
    author = {Berry, Michael Victor and Tabor, M.},
    title = {Level clustering in the regular spectrum},
    journal = {Proceedings of the Royal Society of London. A. Mathematical and Physical Sciences},
    volume = {356},
    number = {1686},
    pages = {375-394},
    year = {1977},
    month = {09},
    issn = {0080-4630},
    doi = {10.1098/rspa.1977.0140},
    url = {https://doi.org/10.1098/rspa.1977.0140},
    eprint = {https://royalsocietypublishing.org/rspa/article-pdf/356/1686/375/62525/rspa.1977.0140.pdf},
}

@article{Bohigas84,
  title = {Characterization of Chaotic Quantum Spectra and Universality of Level Fluctuation Laws},
  author = {Bohigas, O. and Giannoni, M. J. and Schmit, C.},
  journal = {Phys. Rev. Lett.},
  volume = {52},
  issue = {1},
  pages = {1--4},
  numpages = {0},
  year = {1984},
  month = {Jan},
  publisher = {American Physical Society},
  doi = {10.1103/PhysRevLett.52.1},
  url = {https://link.aps.org/doi/10.1103/PhysRevLett.52.1}
}

@article{Huse_2014,
  title = {Phenomenology of fully many-body-localized systems},
  author = {Huse, David A. and Nandkishore, Rahul and Oganesyan, Vadim},
  journal = {Phys. Rev. B},
  volume = {90},
  issue = {17},
  pages = {174202},
  numpages = {5},
  year = {2014},
  month = {Nov},
  publisher = {American Physical Society},
  doi = {10.1103/PhysRevB.90.174202},
  url = {https://link.aps.org/doi/10.1103/PhysRevB.90.174202}
}

@article{Serbyn_2013,
  title = {Local Conservation Laws and the Structure of the Many-Body Localized States},
  author = {Serbyn, Maksym and Papi\ifmmode \acute{c}\else \'{c}\fi{}, Z. and Abanin, Dmitry A.},
  journal = {Phys. Rev. Lett.},
  volume = {111},
  issue = {12},
  pages = {127201},
  numpages = {5},
  year = {2013},
  month = {Sep},
  publisher = {American Physical Society},
  doi = {10.1103/PhysRevLett.111.127201},
  url = {https://link.aps.org/doi/10.1103/PhysRevLett.111.127201}
}

@article{Hsu93,
  title = {Level repulsion in integrable and almost-integrable quantum spin models},
  author = {Hsu, Theodore C. and Angle`s d'Auriac, J. C.},
  journal = {Phys. Rev. B},
  volume = {47},
  issue = {21},
  pages = {14291--14296},
  numpages = {0},
  year = {1993},
  month = {Jun},
  publisher = {American Physical Society},
  doi = {10.1103/PhysRevB.47.14291},
  url = {https://link.aps.org/doi/10.1103/PhysRevB.47.14291}
}

@article{Finkel_Lopez_2015_Yangian_Fibonacci,
title = {Yangian-invariant spin models and {Fibonacci} numbers},
journal = {Annals of Physics},
volume = {361},
pages = {520-547},
year = {2015},
issn = {0003-4916},
doi = {https://doi.org/10.1016/j.aop.2015.07.014},
url = {https://www.sciencedirect.com/science/article/pii/S0003491615002791},
author = {Federico Finkel and Artemio González-López}
}

@article{Finkel_2005,
  title = {Global properties of the spectrum of the {Haldane-Shastry} spin chain},
  author = {Finkel, Federico and Gonz\'alez-L\'opez, Artemio},
  journal = {Phys. Rev. B},
  volume = {72},
  issue = {17},
  pages = {174411},
  numpages = {6},
  year = {2005},
  month = {Nov},
  publisher = {American Physical Society},
  doi = {10.1103/PhysRevB.72.174411},
  url = {https://link.aps.org/doi/10.1103/PhysRevB.72.174411}
}

@article{Potter_2016,
  title = {Symmetry constraints on many-body localization},
  author = {Potter, Andrew C. and Vasseur, Romain},
  journal = {Phys. Rev. B},
  volume = {94},
  issue = {22},
  pages = {224206},
  numpages = {7},
  year = {2016},
  month = {Dec},
  publisher = {American Physical Society},
  doi = {10.1103/PhysRevB.94.224206},
  url = {https://link.aps.org/doi/10.1103/PhysRevB.94.224206}
}

@article{Vasseur_2015,
  title = {Quantum Criticality of Hot Random Spin Chains},
  author = {Vasseur, R. and Potter, A. C. and Parameswaran, S. A.},
  journal = {Phys. Rev. Lett.},
  volume = {114},
  issue = {21},
  pages = {217201},
  numpages = {5},
  year = {2015},
  month = {May},
  publisher = {American Physical Society},
  doi = {10.1103/PhysRevLett.114.217201},
  url = {https://link.aps.org/doi/10.1103/PhysRevLett.114.217201}
}

@book{Mehta1990,
  author    = {Mehta, Madan Lal},
  title     = {Random Matrices},
  publisher = {Academic Press},
  year      = {1990},
  isbn      = {978-0-12-488051-1},
  doi       = {10.1016/C2009-0-22297-5},
  address   = {San Diego},
  copyright = {Copyright © 1990 Elsevier Inc.}
}

@article{FabriciusMcCoy2001,
  author  = {Fabricius, Klaus and McCoy, Barry M.},
  title   = {Bethe's Equation Is Incomplete for the {XXZ} Model at Roots of Unity},
  journal = {Journal of Statistical Physics},
  volume  = {103},
  number  = {5},
  pages   = {647--678},
  year    = {2001},
  doi     = {https://doi.org/10.1023/A:1010380116927}
}

@article{Inozemtsev1990JSP,
  author  = {Inozemtsev, V. I.},
  title   = {On the connection between the one-dimensional {$S {=} 1/2$} {Heisenberg} chain and {Haldane-Shastry} model},
  journal = {Journal of Statistical Physics},
  volume  = {59},
  pages   = {1143--1155},
  year    = {1990},
  doi     = {10.1007/BF01334745},
  received = {1989-07-21},
  month   = jun
}

@incollection{AddarioBerry2008,
  author    = {Addario-Berry, Louigi and Reed, Bruce A.},
  title     = {Ballot Theorems, Old and New},
  booktitle = {Horizons of Combinatorics},
  editor    = {Gy{\H{o}}ri, Ervin and Katona, Gyula O. H. and Lov{\'a}sz, L{\'a}szl{\'o} and S{\'a}gi, G{\'a}bor},
  series    = {Bolyai Society Mathematical Studies},
  volume    = {17},
  pages     = {9--35},
  year      = {2008},
  publisher = {Springer},
  address   = {Berlin, Heidelberg},
  doi       = {10.1007/978-3-540-77200-2_1}
}

@article{Kalmeyer87,
  title = {Equivalence of the resonating-valence-bond and fractional quantum Hall states},
  author = {Kalmeyer, V. and Laughlin, R. B.},
  journal = {Phys. Rev. Lett.},
  volume = {59},
  issue = {18},
  pages = {2095--2098},
  numpages = {0},
  year = {1987},
  month = {Nov},
  publisher = {American Physical Society},
  doi = {10.1103/PhysRevLett.59.2095},
  url = {https://link.aps.org/doi/10.1103/PhysRevLett.59.2095}
}

@article{Zou89,
  title = {Equivalence of fractional quantum Hall and resonating-valence-bond states on a square lattice},
  author = {Zou, Z. and Doucot, B. and Shastry, B. S.},
  journal = {Phys. Rev. B},
  volume = {39},
  issue = {16},
  pages = {11424--11427},
  numpages = {0},
  year = {1989},
  month = {Jun},
  publisher = {American Physical Society},
  doi = {10.1103/PhysRevB.39.11424},
  url = {https://link.aps.org/doi/10.1103/PhysRevB.39.11424}
}

@article{Arovas88,
  title = {Extended Heisenberg models of antiferromagnetism: Analogies to the fractional quantum Hall effect},
  author = {Arovas, Daniel P. and Auerbach, Assa and Haldane, F. D. M.},
  journal = {Phys. Rev. Lett.},
  volume = {60},
  issue = {6},
  pages = {531--534},
  numpages = {0},
  year = {1988},
  month = {Feb},
  publisher = {American Physical Society},
  doi = {10.1103/PhysRevLett.60.531},
  url = {https://link.aps.org/doi/10.1103/PhysRevLett.60.531}
}

@article{Kalmeyer89,
  author = {Kalmeyer, V. and Laughlin, R. B.},
  title = {Theory of the spin liquid state of the Heisenberg antiferromagnet},
  journal = {Physical Review B},
  volume = {39},
  number = {16},
  pages = {11879--11894},
  year = {1989},
  doi = {10.1103/PhysRevB.39.11879},
  url = {https://doi.org/10.1103/PhysRevB.39.11879}
}

@misc{haldane_shastry_repo,
  title ={The raw data supporting this study is available in the GitHub repository: Level-statistics-of-the-disordered-{Haldane}-{Shastry}-model-with-1-r-alpha-interaction},
  url = {https://github.com/Vengat3san/Level-statistics-of-the-disordered-Haldane-Shastry-model-with-1-r-alpha-interaction},
  year = {2026},
}

@article{Moore1991,
title = {Nonabelions in the fractional quantum hall effect},
journal = {Nuclear Physics B},
volume = {360},
number = {2},
pages = {362-396},
year = {1991},
issn = {0550-3213},
doi = {https://doi.org/10.1016/0550-3213(91)90407-O},
url = {https://www.sciencedirect.com/science/article/pii/055032139190407O},
author = {Gregory Moore and Nicholas Read},
}

@article{Quspin1,
	title = {QuSpin: a Python package for dynamics and exact diagonalisation of quantum many body systems part I: spin chains},
	pages = {003},
	author = {Weinberg, Phillip and Bukov, Marin},
	journal = {SciPost Phys.},
	volume = {2},
	year = {2017},
	publisher = {SciPost},
	doi = {10.21468/SciPostPhys.2.1.003},
	url = {https://scipost.org/10.21468/SciPostPhys.2.1.003}
}

@article{Quspin2,
	title = {QuSpin: a Python package for dynamics and exact diagonalisation of quantum many body systems. Part II: bosons, fermions and higher spins},
	pages = {020},
	author = {Weinberg, Phillip and Bukov, Marin},
	journal = {SciPost Phys.},
	volume = {7},
	year = {2019},
	publisher = {SciPost},
	doi = {10.21468/SciPostPhys.7.2.020},
	url = {https://scipost.org/10.21468/SciPostPhys.7.2.020}
}

\end{document}